\documentclass[12pt,preprint]{aastex}

\newcommand{\lyal}{$Ly\alpha$}
\newcommand{\hal}{$H\alpha$}
\newcommand{\hbeta}{$H\beta$}
\newcommand{\lyaha}{$Ly\alpha/H\alpha$}
\newcommand{\lyahb}{$Ly\alpha/H\beta$}
\newcommand{\ebv}{$E(B-V)$}
\newcommand{\vexp}{$v_{exp}$}
\newcommand{\cm}{cm$^{-2}$}
\newcommand{\kms}{km~s$^{-1}$}
\newcommand{\Myr}{M$_\odot\ yr^{-1}$}

\begin{document}

\slugcomment{Accepted by The Astrophysical Journal}


\title{Lyman alpha emission in starbursts:\\ implications for galaxies at
  high redshift\footnotemark}

\author{J.M. Mas-Hesse\altaffilmark{2}}
\affil{Centro de Astrobiolog\'\i a (CSIC--INTA), Madrid,  Spain,
 mm@laeff.esa.es }
\author{D. Kunth}
\affil{Institut d'Astrophysique, Paris, France, 
kunth@iap.fr}
\author{G. Tenorio-Tagle}
\affil{Instituto Nacional de Astrof\'{\i}sica, \'Optica, y Electr\'onica, 
Puebla, Mexico, gtt@inaoep.mx }
\author{C. Leitherer}
\affil{Space Telescope Science Institute, Baltimore, USA, leitherer@stsci.edu}
\author{R.J. Terlevich\altaffilmark{3}}
\affil{Instituto Nacional de Astrof\'{\i}sica, \'Optica, y Electr\'onica, 
Puebla, Mexico, rjt@inaoep.mx}

\and 

\author{E. Terlevich}
\affil{Instituto Nacional de Astrof\'{\i}sica, \'Optica, y Electr\'onica, 
Puebla, Mexico, eterlevi@inaoep.mx}

\footnotetext{Based on observations with the NASA/ESA Hubble
  Space Telescope, obtained at the Space Telescope Science Institute, which
  is operated by the Association of universities for Research in Astronomy,
  Inc., under NASA contract NAS 5-26555}
\altaffiltext{2}{Laboratorio de Astrof\'{\i}sica y F\'{\i}sica Fundamental 
- INTA, Madrid, Spain} 
\altaffiltext{3}{Institute of Astronomy, Cambridge, UK}

\begin{abstract}

We present the results of a high resolution UV 2-D spectroscopic survey of
star forming galaxies observed with HST--STIS.  Our main aim was to map the
\lyal\ profiles to learn about the gas kinematics and its relation with the
escape of \lyal\ photons and to detect extended \lyal\ emission due to
scattering in gaseous halos. We have combined our data with previously
obtained UV spectroscopy on other three star-forming galaxies.  We find
that the P-Cygni profile is spatially extended, smooth and spans 
several kiloparsecs
covering a region much larger than the starburst itself. We propose a
scenario whereby an expanding super-shell is generated by the interaction of
the combined stellar winds and supernova ejecta from the young starbursts,
with an extended low density halo.  The variety of observed \lyal\ profiles
both in our sample and in high redshift starbursts is explained as phases in
the time evolution of the super-shell expanding into the disk and halo of
the host galaxy. The observed shapes, widths and velocities are in
excellent agreement with the super-shell scenario predictions and represent
a time sequence.  We confirm that among the many intrinsic parameters of a
star forming region that can affect the properties of the 
observed \lyal\ profiles,
velocity and density distributions of neutral gas along the line of
sight are by far the dominant ones, while the amount of dust will
determine the intensity of the emission line, if any.

\end{abstract}



\section{Introduction}

Galaxies with ongoing star formation display characteristic emission lines
whose strength  often dominates the appearance of the optical spectrum
\citep{kennicutt}. The ionizing radiation from newly formed stars and its
interaction with the surrounding gas generate collisionally excited and
recombination lines which become detectable at the highest observable
redshifts \citep{melnick}. Model spectra of young
populations predict \lyal\ to be the strongest emission line in the
optical/near-infrared (IR) spectral range for the simplified assumption of
Case B recombination and low metal content (see \citet{schaerer} for a very
recent set of model calculations). Therefore \citet{partridge}
suggested the \lyal\ line as an important spectral signature in young
galaxies at high redshift as the expected \lyal\ luminosity could amount to a
few percent of the total galaxy luminosity.

Typical \lyal\ fluxes of $10^{-15}$~erg~s$^{-1}$~cm$^{-2}$\ are
expected for galaxies at redshifts around 3 with star-formation rates of
order $10^2$~\Myr. Such values have been within the reach of even
relatively modest-sized instruments for several decades. Major
observational efforts were undertaken to search for \lyal\ emission from
such galaxies \citep{djorgovski}. Although quite a few \lyal\
emitters powered by starbursts have been found (e.g., \citet{kudritzki}, 
\citet{rho00}, their numbers are generally much lower than
expected from the observed star-formation rates and Case B recombination
conditions. 

The assumption of the \lyal\ intensity as produced by pure recombination in
a gaseous medium may be too simple. \citet{meier81}, \citet{hartmann}, 
\citet{neufeld}, and \citet{charlot} considered the
effects of dust on \lyal. \lyal\ photons experience a
large number of resonant scatterings in neutral atomic hydrogen, thereby
increasing the path length and the likelihood of dust scattering and
absorption. This process can be very efficient in removing \lyal\ photons
from the line of sight to the observer, leading to much lower line
strengths in comparison with the idealized Case B. Depending on the aspect
angle of the galaxy as seen from the observer, this may lead to a decrease
of the \lyal\ equivalent width. On the other hand, \lyal\ may actually be
enhanced due to the presence of many supernova remnants which form during
the starburst \citep{shull}. The net result is
controversial. \citet{bithell} finds supernova remnants to be an important
contributor to the \lyal\ strength whereas \citet{charlot} reach the
opposite conclusion.

The theoretical situation is sufficiently complex that observational tests
are required. The most obvious test are measurements of \lyal\ in local
starburst galaxies whose redshifts are sufficiently large to permit
observations of their intrinsic \lyal\ outside the geocoronal and Galactic
interstellar \lyal. Observations of local starbursts have indeed been
performed with the {\it IUE} satellite (\citet{meier81};
\citet{hartmann}; \citet{calzetti}; \citet{terlevich93}; \citet{valls93}).
Again, the results are controversial. For instance, Calzetti \& Kinney and
Valls-Gabaud find \lyal\ strengths in agreement with pure recombination
theory whereas Hartmann et al. and Terlevich et al. conclude that
significant dust trapping of \lyal\ photons must occur.

The superior spectral and spatial resolution of HST's ultraviolet (UV)
spectrographs has allowed new insight into the formation process of
\lyal. HST--GHRS spectroscopy of eight gas-rich irregular galaxies by
\citet{klyal98} indicates yet another, and most likely the dominant
parameter governing \lyal\ emission: neutral gas kinematics. Kunth et
al. found \lyal\ emission with blueshifted absorption in four of the
galaxies. In these objects the OI and SiII absorption lines are also
blueshifted, suggesting an outflow of the neutral gas with velocities
around  200~\kms. The other four galaxies show broad damped \lyal\ absorption
profiles centered on the wavelength of the ionized gas with no detection of
\lyal\ emission. The observed galaxies span a metallicity range of more
than a factor of 10 and display no correlation between metal abundance and
\lyal\ emission strength, a correlation that had been postulated on the
basis of the IUE spectra and on theoretical grounds due to the appearance
of dust. The velocity structure of the neutral gas in these galaxies is the
driving factor that determines the detectability of \lyal\ in
emission. When most of the neutral gas is velocity-shifted relative to the
ionized regions, the \lyal\ photons can escape, a suggestion supported by
recent models of \citet{gtt99}.  
 Nevertheless, while the velocity structure of the neutral gas is the
  driver for the detectability of the \lyal\ emission, we want to stress
  that the amount of dust will still be responsible for the {\em intensity}
  of the line, when observed. The \lyal\ photons not affected by 
  resonant scattering will still be strongly affected by dust
  extinction, which has a maximum in the UV range, around the \lyal\
  wavelength. The properties of the neutral gas (density, kinematics,
  covering factor) determine thus the shape of the profile and the
  equivalent width of the line, while the amount of dust drives only its
  intensity.   
The implication is that feedback from the
massive stars via ionization and the creation of superbubbles and
galactic-scale outflows lead to the large variety of \lyal\ profiles. The
escape of \lyal\ photons depends critically on the column density of the
neutral gas and dust, the morphology of the supershells, and the kinematics
of the medium.  Since these effects can be highly stochastic, theoretical
predictions for the \lyal\ strength are quite uncertain, and empirical
guidelines are called for.

A similar pattern seems to apply to high redshift galaxies. At $2.5<z<
5.2$, about half of the galaxies found show \lyal\ in emission 
(\citet{steidel}; \citet{rho00}; \citet{frye02}). At
the same time, the \lyal\ line is asymmetric, displaying an extension
towards larger wavelengths. The blue ``edge'' of the line could be
described as showing a P-Cygni profile, and the centroid of the line is
redshifted by some hundred \kms\ with respect to the metallic
absorption lines. This is consistent with gas outflows, breakout of the gas
bubble produced by the star forming region, and results in the 
escape of \lyal\ emission. 

In order to analyze the spatial (kinematical) structure of the \lyal\
emission in nearby star forming galaxies, we used the Space Telescope
Imaging Spectrograph (STIS) on board HST to reobserve three galaxies of the
\citet{klyal98} sample, two of which clearly showed with GHRS a P-Cygni
\lyal\ profile (Haro~2 and IRAS~0833+6517) and a third one that only
presented a broad \lyal\ absorption (IZw18). We aimed also to analyze
whether \lyal\ could be leaking in external regions, while being completely
absorbed around the starburst itself.

The data and data analysis technique are presented in Section
2; the expected structure of an HII region and subsequent effects on the
resulting \lyal\ profiles follow in Section 3; the superbubble generated by
the HII region and their time evolution are analyzed in Section 4, together
with the expected effect on the observed \lyal\ line profile. The general
discussion is presented in Section 5, and the particular implications for
high redshift galaxies, in Section 6. Finally, Section 7 of the paper
presents the summary and conclusions of our work.

\section{HST--STIS observations} 

\placetable{sample1}

We list in Table~\ref{sample1} the basic data for the galaxies observed
with HST--STIS. The journal of observations is summarized in
Table~\ref{observations1}. We will complement our discussion with some data
presented by \citet{thi97}. These autors analyzed a sample of 3 compact
galaxies experiencing strong starbursts. They found \lyal\ emission in only
one case, Tol~1214-277. The line was very strong and rather symmetric with no
 evidence of blueshifted absorption. They concluded that in this
case the line was visible
because the covering factor of the neutral cloud was small, leaving
 \lyal\ photons  able to escape through paths relatively free of
HI gas.

\placetable{observations1}

Previous GHRS spectra were taken through the Large Science Aperture (LSA),
comprising roughly 1\farcs7 $\times$ 1\farcs7 (2\arcsec$\times$2\arcsec for
the pre-Costar IZw18 data), centered on the maximum UV continuum. STIS data
have been obtained through the 52\arcsec$\times$0\farcs5 slit on the FUV
MAMA and optical CCD detectors. 

In the FUV range the G140M grating was used.  The data have been extracted
with standard STSDAS procedures. The different integrations obtained were
combined and averaged with the MSCOMBINE task to improve the signal to
noise ratio. The detector and geocoronal background was removed averaging
100 rows on areas below and above the stellar continuum. While the detector
background was negligible for Haro~2 and IRAS~0833+6517, it affected
significantly the bluest continuum of IZw18.

Low resolution optical spectra were taken with the STIS CCD using the G430L
grating, in order to compare locally the extension and intensity of \lyal\
and optical emission lines, namely [OII], [OIII] and \hbeta. The optical
spectrum of IZw18 was underexposed, but we could clearly detect the
extension of the optical lines for Haro~2 and IRAS~0833+6517.

The UV spectral images have a spatial scale of 0\farcs029 and a spectral
dispersion of 0.053 \AA/pix, giving a spectral resolution around 0.15 \AA,
translating to $\sim$ 37 \kms\ at the \lyal\ wavelength. 
The optical images have correspondingly scales
of 0\farcs05 and spectral dispersion of 2.7 \AA/pix (spectral resolution
$\sim$  6.7 \AA).  We have multiplied
the resulting fluxes by the STIS parameter {\tt diff2pt} (diffuse to point
source conversion factor for absolute photometry), as required for
non-extended sources - although resolved, the objects are indeed rather
compact.

In the next section we describe in detail the observations object by
object.

\subsection{Haro~2} 

We show in Fig.~\ref{2dh2} the STIS spectral image of
Haro~2 with different scales. The slit is oriented along the minor axis of
the galaxy (during a first observation with the slit along the major axis
the aperture was unfortunately misplaced due to an observer error). It can
be seen that the UV continuum is very compact, extending over only around
0\farcs9 (corresponding to $\sim 85$ pc).  There is a very strong, quite
extended and spatially asymmetric \lyal\ emission line. \lyal\ emission is
detected over ~7\arcsec, ~6\arcsec\ of which are located to the upper (SW)
part of the slit. There are hints that the nebulosity is even more
extended, at a fainter level, to this direction.  The peak of the \lyal\
emission is slightly offset (by 0\farcs2 or 19 pc) from the peak of the UV
continuum. A spatial profile of the \lyal\ emission and the UV continuum is
shown in Fig.~\ref{h2lyalspat}. In Fig.~\ref{h2lya-1} we display the extracted
spectrum of the central region of this galaxy.

The most striking result from these data is that the conspicuous and completely
black absorption edge affects the blue wing of the \lyal\ profile 
over the whole region where \lyal\ is detected, at essentially the same
velocity. This implies that the neutral gas that causes the resonant
scattering is approaching us at around
200~\kms\ and  extends over at least $\sim$8\arcsec ($\sim$750 pc) with no
 hint of any velocity structure. We show in Fig~\ref{h2lya} the \lyal\
profile at different positions over the slit. The zero in the velocity
scale corresponds to the \lyal\ centered at the redshift of the central HII
region determined from the optical emission lines. These plots confirm that
the blue edge of the \lyal\ profile always appears at basically the same
velocity, with variations smaller than $\pm$ 30 \kms. Note as well that the
red wing of the \lyal\ emission profile extends  essentially to the same
velocity ($\sim + 500$~\kms) all along the slit, independently of the peak
intensity of the line. As we will discuss later the
emission from this wing originates from a receding shell far from the 
central HII region.

\placetable{observations2}

The optical spectral image is shown in Fig.~\ref{h2-o2d}. The optical
continuum, from 3000 to 5000 \AA is very blue and appears to be dominated
by very young stars, as illustrated in Fig.~\ref{h2hb}. The already known
Wolf-Rayet feature at around 4686 \AA\ is clearly detected in the
spectrum.  Emission lines are detected  over
$\sim$1\farcs5 extent, which corresponds to the core of the \lyal\ line. The
optical lines show some asymmetry at the upper part of the slit. The
spectrum has enough signal-to-noise so as to measure the total flux of
\hbeta\ and the OII+OIII lines; these are listed in
Table~\ref{observations2}, together with the \lyal\ flux.  It is important
to note that these fluxes are integrated over the same area
(1\farcs5$\times$0\farcs5) hence their ratios reflect the intrinsic local
ratios in the gas. The observed \lyahb\ ratio is only $\sim$2.1, much below
the theoretical value \lyahb$\sim$33. \citet{mhk99} measured an extinction
of \ebv=0.22, as derived from the Balmer lines observed through a rather
large aperture containing the whole ionized region. With this extinction,
the expected \lyahb\ ratio would be around 3.4 (assuming a Large Magellanic
Cloud (LMC) extinction law in the UV - see \citet{mhk99}). This implies
that the effect of resonant scattering has lowered the \lyal\ intensity by
at least 40\%. Nevertheless, we stress that this extinction value
represents an average over a large region. Since the Balmer line ratio is
not measured with the same spectral resolution as the \lyal\ line, it is not
possible to disentangle effects due to dust extinction with those of resonant
scattering.

This galaxy exemplifies the paradigm defining the detectability of the
  \lyal\ emission line: while most of the line emission has been destroyed
  in this case by dust absorption, the line is still detectable, 
  and indeed quite prominently,
  due to the  kinematical configuration of the neutral gas
  surrounding the HII region.  If there were no dust at all, the line would
  remain undetectable unless the neutral gas were expanding. On the other
  hand, if there were not neutral gas at all, both the \lyal\ emission line
  and the surrounding continuum would experience roughly the same
  absorption, so that the equivalent width of the line would remain
  unaffected, independently of the amount of dust. Only the absolute line 
intensity would be strongly dependent on the presence of dust.

The evolutionary synthesis modelling by \citet{mhk99} yielded an age for
the starburst episode around 4.8 Myr and more than $6.6\times10^6$
M$_\odot$ of gas having been transformed into stars. As cited by these
authors, there is evidence in Haro~2 of previous recent starburst episodes,
which could have taken place in the last 50 Myr. 

\subsection{IRAS~0833+6517} 

The STIS spectral images of IRAS~0833+6517 reveal a much more complex
structure than Haro~2 (see Fig.~\ref{2dir}).  The UV
continuum shows a patchy distribution, and is extended over $\sim$3\arcsec\
(1.2~kpc) -- its limits are diffuse: the UV continuum seems to be even more
extended at a lower brightness level.  There is also a strong, quite
extended and spatially asymmetric \lyal\ emission line. \lyal\ emission is
detected over at least $\sim$10\arcsec\ (4.1~kpc). \lyal\ emission shows 2
main peaks: the brightest is slightly offset with respect to the center of
the UV emission, $\sim$0\farcs3 (124~pc) to the upper (S) part of the
slit. The second peak is also offset, $\sim$1\farcs2 (496~pc) to the lower
(N) part of the UV continuum. Some weaker peaks can also be identified on
the image. This is shown in Fig.~\ref{irlyalspat}, where we plot the
spatial profile of the \lyal\ emission, on top of the continuum profile.
The extracted spectrum in the \lyal\ region is given in Fig.~\ref{irlya},
showing the asymmetric \lyal\ emission line with a broad absorption to the
blue.

We show in Fig.~\ref{irlya-2} the \lyal\ emission profile across different
regions.  As already found in Haro 2, the blue absorption edge has no 
 any significant velocity structure over at least the 10\arcsec\ (4.1
kpc) along the slit where emission is detected (the velocity at which the
profiles go to zero in the different regions is within the range $-50$ --
$+50$~\kms. Here again, the neutral gas in front of the HII region, within
at leat 4.1 kpc, is approaching us at basically the same velocity,
$\sim$300 \kms. Moreover, the red wing of the emission profile extends 
 to roughly the same velocity, around 700 \kms, as also found on
Haro~2.  

We confirm the detection made on the GHRS spectrum of a secondary \lyal\
emission line, blueshifted by around $-$300 \kms\ with respect to the HII
region velocity determined from the optical emission lines. This secondary
emission shows some interesting features:

\begin{itemize} 

\item It is clearly offset from the continuum, peaking at $\sim$0\farcs9
(370~pc) from the UV continuum, to the upper (S) part of the slit.  The
emission, at a weaker level, extends clearly to the lower (N) part of the slit,
at least over the region where there is some UV continuum.

\item Although the signal to noise is low, this secondary emission peak
shows some velocity structure, with components redshifted by
$\sim$100 \kms\ from its average centroid.  In general, the profile is not
a gaussian one and seems to be the convolution of various emissions at
different velocities.

\end{itemize} 

We show in Fig.~\ref{h2-o2d} the spectral image of IRAS~0833+6517 taken in
the optical range.  The continuum has a narrow peak about
$\sim$0\farcs3 wide (124~pc), on top of a weaker extension over
$\sim$2\arcsec (826~pc). There is a region to the upper part of the slit
which is clearly more conspicuous in the UV than in the optical. It is here
where \lyal\ and the forbidden oxygen lines peak. This region is offset by
$\sim$1\farcs2 (496~pc) from the optical continuum maximum. We show in
Fig.~\ref{irhb} the optical spectra of the bright nucleus and of this
region.

We believe that the maximum  of the optical continuum  is due to relatively 
older stars, with a weak contribution to the UV and to the ionization.  Indeed,
coincident with the optical continuum maximum there is a local minimum in
all emission lines strengths.  The optical continuum shows a very prominent
Balmer decrement, as well as Balmer absorption lines of stellar origin, but
only in the central region (see Fig.~\ref{irhb}). The optical continuum in
the upper part, where the emission lines peak, is much bluer and with
weaker Balmer decrement and Balmer stellar absorptions, as shown on the
figure. \citet{rosa98} presented a detailed analysis of the massive young
stellar population in this galaxy based on evolutionary synthesis models.
They derived a relatively old age for the burst between around 6 and 7
Myr. These authors concluded that there was a significant dilution by an
underlying older stellar population, which we identify indeed with the
population producing the maximum of the optical continuum.  

Finally, [OII], [OIII] and \hbeta\ are detected over $\sim$3\farcs5 (1.4~kpc)
and coincide again with the core of the \lyal\ emission. The measured \lyahb\
ratio is $\sim$2.1, indeed similar to the value measured in Haro~2 (see
Table~\ref{observations2}). \citet{rosa98} quote a large range of
extinction values for this galaxy, with \ebv\ between 0.17 and
0.52. According to these authors, the value derived from the Balmer lines
ratio is \ebv\ = 0.52. Assuming this extinction and an LMC law, the
expected \lyahb\ ratio would be only around 0.28, much smaller than the
value measured by us. On the other hand, if we consider the extinction
value derived from the continuum fits, with \ebv\ = 0.17, and the same LMC
law, the expected ratio would be much higher, \lyahb\ $\sim$ 7, indeed
significantly higher than the measured value. We conclude again that
although the effects of extinction and scattering can not be disentangled,
our results show clearly that both effects must significantly affect the
observed properties of the \lyal\ emission line.

\subsection{IZw18}

We show in Fig.~\ref{2diz} the STIS UV spectral image of IZw18.  The UV
continuum is quite extended (over $\sim$2\arcsec, corresponding to
$\sim$97 pc). The total flux within the STIS aperture is around a factor
4 lower than within the GHRS aperture (2\arcsec$\times$2\arcsec,
pre-Costar data), which is consistent considering the  extension of
the source.

A very broad and damped \lyal\ absorption is detected, with complete
blackening at the center of the line, as shown in Fig~\ref{izlya}.  The
longer wavelength range of STIS allows for a more complete coverage of the
absorption red wing, as compared to the GHRS data from \citet{klyal98},
confirming the properties of the damped absorption profile reported there.
No emission is seen at any position along the slit and regardless the
intensity of the continuum. This is clearly understood if the neutral gas
covers the whole ionized region with a large column density. This neutral
gas must be static with respect to the central HII region all along the
region covered by the slit, so that resonant scattering affects all 
\lyal\ photon emitted by the central HII region. The complete blackening of
the absorption profile indicates that essentially all photons have been finally
destroyed, most likely by dust absorption.

\section{The structure of an HII region and effects on the resulting 
\lyal\  profiles}

An HII region surrounding a cluster of massive newly formed stars should
produce an intrinsic \lyal\  emission line with a \lyaha\ ratio 
around 11 and an intrinsic velocity dispersion ($\sigma$) similar to the
 optical Balmer lines one. Nevertheless, there are several factors
affecting significantly both the intensity and the profile of the \lyal\ 
line as observed from outside the HII region. We will discuss the different
effects in this section aiming to explain the observed \lyal\  profiles.

The main effects contributing to modify the \lyal\  emission line are the
following:

\begin{itemize}

\item Absorption by dust. The known extinction curves peak in the far UV
range (around 1000~\AA), so that the interstellar extinction will be
maximum around \lyal. As an example, an \ebv\ of 0.2 will yield the
absorption of 38\% \hal\ luminosity, but of 96\% of the \lyal\  
emission -- according to the Small Magellanic Cloud law, generally valid 
in these environments \citep{mhk99}.

As a first approximation, the \lyal\ line and the surrounding continuum
should experience the same extinction by dust (assuming there is no neutral
gas along the line of sight), so that the equivalent width of the line would
not be affected. Therefore, in regions where the continuum is well
detected, dust absorption alone could not explain the weakness or even
absence of \lyal\ in emission. 

Nevertheless, we want to remind that the spatial distribution of stellar
continuum sources, ionized gas and dust clouds might be very different, as
found in nearby starburst galaxies (see for example the analysis of
NGC~4214 by \citet{maiz}). This spatial decoupling might lead to very
different extinctions affecting the UV continuum and the emission lines, as
discussed by \citet{mhk99}.  
  
\item Scattering by neutral hydrogen. \lyal\ photons travelling through
neutral hydrogen will suffer resonant scattering which will re-distribute
them within the cloud. This effect dramatically increases  the sensitivity
of \lyal\  photons to dust absorption.  In the presence of neutral hydrogen
we therefore expect the complete destruction of \lyal\  photons by dust
absorption, even in environments with relatively low dust abundance.

\item Presence of expanding shells.  The presence of an expanding shell
around the HII region can dramatically affect the shape and intensity of the
\lyal\  emission line. We have identified 4 major elements: 

\begin{itemize} 

\item If the expanding shell is sweeping the neutral gas, the resonant
scattering will affect photons with energy slightly higher than
those of the central \lyal\  emission line. As a result, only part of the
\lyal\  emission will be able to go through the neutral medium and become
visible.  P-Cygni like profiles will be expected in these cases, as we
will show later.

\item  If a fraction of ionizing photons escapes the central HII region, 
the ionization front could reach the internal layers of the shell.
We would expect to detect then a redshifted \lyal\  emission component 
originated at the inner surface of the receding shell. 

\item In addition, part of the \lyal\  photons produced in the central HII
region could be backscattered by the neutral layers of the 
receding shell to the obsever's line
of sight, appearing also redshifted (with respect to the HII region
systemic velocity) by the shell expansion velocity.

\item Finally, a recombining region, behind the leading shock, could
cause a secondary \lyal\  emission component, which would be
observed blueshifted by the shell expansion
velocity. Since this component originates {\em outside} the neutral
shell, it
wouldn't be affected by resonant scattering by the neutral layers.

\end{itemize}
\end{itemize} 

We will analyze in the next sections in more details effects
related to resonant scattering and the presence of an
expanding shell.

\subsection{Effect caused by resonant scattering}

Neutral hydrogen will scatter resonantly the \lyal\ photons
inside the cloud. In Fig.~\ref{voigt} we show the expected absorption
profiles produced by a layer of neutral hydrogen with different column
densities. Profiles have been computed using the XVoigt code
\citep{mar95}, assuming a thermal broadening of the neutral cloud with $b =
20$ \kms. The Figure shows how huge are  the effects of resonant scattering
by neutral hydrogen: for relatively small column densities ($N
\sim 10^{14}$ \cm) the center of the line reaches total absorption. For
column densities higher than $N \sim 10^{18}$ \cm, the line becomes damped
and the full saturation (hence no transmission) spreads rapidly towards
both sides. Remark from the Figure that for $N \sim 10^{21}$
\cm\ all photons emitted by particles within a range in velocity of
 $\pm$ 1000 \kms\ will be completely absorbed. This  not
only affects the \lyal\ emission line photons, but also those photons from the
surrounding continuum emitted by the central cluster of massive
stars. Moreover, the wings of the absorption would affect photons at
energies corresponding to $\pm$ 4000 \kms\ from the line center.

In principle resonant scattering by neutral hydrogen does not destroy
 the incident \lyal\  photons. In a completely dust-free cloud, 
photons would be internally  scattered  until they reach the external 
surface of the cloud from where they are  able to escape. 
We would expect \lyal\  photons
to leak from the overall surface of the neutral cloud surrounding
the HII region, producing a very low surface brightness. 
\citet{all1,all2} have  discussed  the effects of \lyal\ photons
  scattering by neutral Hydrogen  in a dustless and static medium. They
   assume that no photon  become destroyed  (in absence of
  dust) and then analyze  the redistribution in frequencies of the  escaping 
radiation. They predict the formation of a rather narrow
  absorption trough at the line center, with extended red and blue
  wings but their profiles differ significantly from the observed
  ones basically because the effect of
  dust was not considered in these works.  Indeed  \citet{klyal98} showed 
that when a static cloud of neutral gas is
  surrounding the HII region the resulting profile corresponds to a
  typical Voigt-like absorption line. 
We stress again that even very small amounts of dust
 completely destroy all scattered photons via
multiple scattering events. Our simulations consider that all photons
 affected by scattering, according to the corresponding Voigt function, 
will be finally absorbed by dust and will not be re-emitted. 

\subsection{Effects caused by an expanding shell} 

According to the profiles shown in Fig.~\ref{voigt} it would
become nearly impossible to detect \lyal\  in emission from starburst
galaxies, since they are usually immersed in rather dense neutral
clouds. However, we argue that the presence of an expanding shell, with
properties evolving concurrently with 
the central cluster of massive stars, allows for an explanation 
of the various profiles actually detected in most objects, as
discussed in \citet{gtt99}. In the following discussion the expansion
velocity of the shell, \vexp, will refer to the velocity of the shell with
respect to the central static HII region. 

We first show in Fig.~\ref{prof1} the expected \lyal\  profiles when the
neutral gas surrounding the HII region is expanding at a certain velocity
(\vexp), for different HI column densities along the line of sight. As
explained above, resonant scattering in the expanding neutral gas 
affects photons with higher energies than those emitted in the \lyal\  line by
the central HII region. As a result, the absorption of \lyal\  photons 
mostly affects  the blue wing of the emission profile producing a classical
P-Cygni shape. The damped part of the absorption profile remains
completely black and the emission line becomes strongly asymmetric. If the
HI column density is high enough (in our example, above around $10^{21}$
\cm), the absorption is total and only a broad damped profile would be
detected. The resulting spectrum will nevertheless depend on the
convolution of different parameters: expanding velocity of the shell, HI
column density along the line of sight, intrinsic intensity and width of
the central \lyal\  emission line, etc... so that we  expect to detect the
different cases presented in Fig.~\ref{prof1} almost randomly. 

Note that  P-Cygni  \lyal\ emission profiles
show a peak which appears redshifted with respect to the intrinsic line by
up to several hundred \kms.  This is shown as an example in
Fig.~\ref{prof2}. The broader the intrinsic \lyal\ emission line, the
redder the peak can appear after partial absorption by an expanding
shell. Clearly, the effect of resonant scattering within a dense, expanding
neutral shell leading to a redshifted \lyal\ emission peak, should not be
taken as evidence for the presence of outflowing ionized gas. 

In the presence of a neutral expanding shell we expect to detect 
interstellar absorption lines associated to the neutral or weakly ionized
gas. These will appear
blueshifted with respect to the systemic velocity by $-$\vexp, as indeed
found by \citet{klyal98}. 

If the expanding shell remains symmetric, the internal surface of
the receding section of the shell will also  affect the
observed \lyal\ profile: a fraction of the ionizing photons
can ionize this internal surface of the shell. The photons produced by the
approaching section of the shell would be immediately scattered by the
neutral gas. Since both the ionized and neutral layers are expected to
expand at the same velocity, the absorption would be very efficient and no
photon can escape. On the other hand  photons originated
at the receding section of the shell would be emitted with a redshift
corresponding to $+$\vexp and will be able to travel 
freely through the approaching neutral layers of the expanding shell.

Moreover, a fraction of the photons scattered by the neutral layers on the
receding shell will end being re-emitted on the direction of the line of
sight, also with a redshift of $+$\vexp, and  able to go through
the neutral layers of the approaching shell. As a result, we expect
to detect an additional \lyal\ emission component peaking at $+$\vexp. This
component should be broader than the intrinsic \lyal\ line, since the
resonant scattering affects photons over a rather wide range of
energies. Nevertheless, we expect this component to be rather weak (at a
level of few percent of the intrinsic \lyal\ line)  producing just a
broadening at the red wing of the resulting \lyal\ emission profile since
 most of the photons affected by scattering would be destroyed by dust.  This
effect is shown in Fig.~\ref{bc1}, where we have plotted the expected
resulting profile assuming a component originated at the receding shell
(both by photoionization and backscattering) amounting to 10\% of the
intrinsic \lyal\ emission line.

 \citet{all3} have extended their models \citet{all1,all2} to a
 spherical, expanding supershell. As discussed in
  Sect.~3.1, these models do not consider the effects of dust absorption,
  and only  allow for multiple backscattering events on the internal layers of
  the expanding neutral shell, both on the approaching and on the receding
  sides. After each backscattering  photons
   get an additional redshift of $+$\vexp\ but since no
  photon destruction is considered several redshifted
  emission peaks are produced by multiple backscattering. However we argue
  that when
  absorption by dust is taken into account,  multiple peaks anticipated
  by these authors would disappear. Moreover, the intensity of the
  single redshifted peak would be much weaker than their prediction since
  most photons affected by scattering are destroyed before
  being back-emitted. We have not performed a line transfer model  to
  reproduce this effect. We  just qualitatively show that both 
  backscattered photons and the \lyal\ ones originating at the internal,
  ionized layers of the expanding shell,
 will produce an additional weak
  and broad emission component centered at around $+$\vexp.

Finally, an ionized region can develop at the external shock front of the
shell. The recombining medium immediately behind the shock front would
produce an additional \lyal\ emission line unaffected by resonant
scattering.  This emission line observed at a blueshift
corresponding to the expansion velocity of the shell, $-$\vexp, 
if present,  would fill partially the damped absorption profile. Its
redshifted counterpart, on the other hand, will be scattered and fully
absorbed within the large column density of neutral matter in the receding
part of the shell.

The above discussion aims to confront the reader with the view that there
are many parameters related to the structure of the HII region that affect
the resulting \lyal\  profile. In particular, the velocity and density
structure of the neutral gas along the line of sight seems to be a
dominant factor for the visibility of the line.

\section{\lyal\  emission in starburst galaxies: an evolutionary view}

In realistic cases, bubbles and superbubbles generated around an HII region
will evolve with time, in parallel with the evolution of the massive stars
themselves. As a result, the physical conditions that the \lyal\  photons 
experience when escaping from the HII region will differ
drastically. Therefore the expected \lyal\  profiles will be strongly
dependent on the geometry and evolutionary state of the starburst. 

We show in Fig.~\ref{model1} the predicted evolution of the expanding
shells generated by an HII region as they interact with the disk and the
halo of the host galaxy.  The figure makes use of the results from the
numerical calculations presented in \citet{gtt99}, and illustrates in a
simplified way the different elements that determine the properties of the
\lyal\ emission profile. We have identified 6 basic steps, as follows:

\begin{itemize} 

\item[i)] Initially, when a star-forming episode starts, a central HII
region begins to develop. At this phase, if the neutral gas surrounding the
starburst region has HI column densities above 10$^{14-15}$ \cm, an
absorption line centered at the systemic velocity of the galaxy will be
visible, independently on the viewing angle. If the total HI column density
along the line of sight is higher than around 10$^{18}$ \cm, a damped \lyal\ 
absorption profile will be detectable. It is important to stress that
during this early phase  the Balmer lines will be strongest,
due to the high ionizing flux produced by the most massive stars.

\item[ii)] The situation changes drastically and becomes a strong function
of viewing angle, once the mechanical energy released by the starburst is
able to drive a shell of swept up matter to exceed the dimensions of the
central disk. Then, upon the acceleration that this shell experiences as it
enters the low density halo, it becomes Rayleigh--Taylor unstable and
fragments.  This event allows the hot gas (composed basically by matter
recently processed by the starburst), to stream with its sound speed
between fragments and follow the shock which now begins to form a new shell
of swept--up halo matter. Another consequence of blowout is the fact
that the ionizing photons from the recent starburst are now able to
penetrate into the low density halo, and manage to produce an extended
conical HII region that reaches the outskirts of the galaxy. Given the low
densities in the halo it is likely that this matter will remain ionized for
a time that well exceeds the duration of the starburst activity. This also
implies that some UV photons are at this stage able to stream freely into
the inter galactic medium. The predicted expansion speed of this second
shell formed in the halo would be around several hundred \kms. 

An observer looking then at the starburst through the conical HII region
will be able to detect the strong \lyal\  emission line produced by the
central HII region, centered at the systemic velocity of the galaxy.  On
the other hand, an observer looking outside the conical HII region
will still detect a broad absorption profile at any evolutionary state.

\item[iii)] Sooner or later, recombination will begin to be frequent enough
in the expanding shell. This will cause a strong depletion of the ionizing
radiation which formerly was able to escape the galaxy after crossing the
extended conical HII region. Recombination in the expanding shell will
produce an additional \lyal\  component, which the observer will detect
blueshifted according to the expansion velocity of the shell, $-$\vexp.
Clearly, if the shell were symmetrical and one could see the emission
coming from the whole of it, a top-hat line profile would have to be added
to the central HII region \lyal\  emission line. If only a fraction of the
shell is observed, we would expect to see 2 additional emission peaks centered 
at $\pm$\vexp.

\item[iv)] There are three efects that eventually lead to the trapping of
the ionization front within the expanding shell.

\begin{itemize}

\item The first one is the increasingly larger amount of matter swept into
the expanding shell, as this ploughs into the halo. 

\item The growth of the shell dimensions also implies less UV photons
impinging, per unit area and per unit time, at the inner edge of the shell.

\item Finally, in the case of a nearly instantaneous starburst, the
production of UV photons starts to decrease drastically (as $t^{-5}$) after
the first 3.5 Myr of evolution.  

\end{itemize} 

The trapping of the ionization front will yield the formation of a neutral
layer at the external side of the expanding shell.  All these effects lead
then to an increasingly larger  saturated absorption, as the external
neutral layer will resonantly scatter the \lyal\  photons. As discussed
above, this absorption will be blueshifted with respect to the \lyal\ photons
emitted by the central HII region by $-$\vexp, leading to the formation
of a P-Cygni profile with a fraction of the intrinsic \lyal\ 
emission beeing absorbed.   

In addition, the profile will be contributed by the \lyal\  radiation arising
from the receding section of the shell, both by recombination on the
ionized layer, and by backscattering of the central \lyal\  photons by the
neutral layer. Since this contribution will be redshifted by 2$\times
v_{exp}$ with respect to the absorbing layer, it will be essentially free
from resonant scattering and will be able to escape the region almost
unaffected. 

Under such circunstances the resulting \lyal\  emission line will
be strongly depleted, showing a very asymmetric profile with a sharp blue
edge and a line centroid shifted towards the red at a
$\lambda$ different from the rest velocity of the host galaxy. The red wing
of this profile will be furthermore broadened by the contribution from the
receding shell.

\item[v)] Under some conditions (for specific values of the shock
  velocity $V_s$ and halo density $n_{halo}$), 
the leading shock front may become radiative. 
An additional \lyal\ 
component would thus arise as the shocked gas undergoes recombination. This
should happen after a cooling time ($t_{\Lambda} = 1.5 k T/(4 n_{halo}
{\Lambda}$); where $T = 1.4 \times 10^7 (V_S/1000 km s^{-1})^2$, $\Lambda$
is the cooling rate and $k$ the Boltzmann constant).  
Thus a combination of shock speed and
density of the background halo may lead to a $t_{\Lambda}$ smaller than the
dynamical time and at that moment \lyal\  emission will be produced as
the shocked gas recombines. This additional \lyal\  emission component would
appear blueshifted at $-$\vexp, and wouldn't be absorbed being
ahead of the HI layers.

\item[vi)] At a later phase, the shell will be  completely recombined
and  have already substantially slowed down its expansion. \lyal\ will
show a broad absorption with only a small blue shift.  The recombination
lines from the HII region will be very weak.

\end{itemize}

The scenario here presented assumes a non-steady state of the star
formation process. It implies therefore either a (nearly) instantaneous
starburst or the first phases of a star formation episode extended in
time. In the case of a continuous star formation regime, a steady state
where the birth and death of massive stars balances, so that the ionizing
flux remains basically constant, is achieved only after the first 15 Myr of
evolution \citep{cmh94}. If the process is active for much longer periods
of time, the shell will finally blowout in the intergalactic medium, and
no \lyal\ emission lines with P-Cygni profile would be expected,
but pure emission or absorption, depending on the orientation.

\section{General discussion}

Fig.~\ref{examples} shows three GHRS low resolution examples corresponding
to typical cases as described above. T1214-277 shows a prominent and
symmetric \lyal\ emission line. The large equivalent width of the line
($\sim 90$ \AA) points to a very young starburst. We postulate that this
object would correspond to steps ii--iii above. In IRAS~0833+6517 the
ionization front has already been trapped, so that a neutral layer has
already developed, giving rise to a blueshifted absorption and the
corresponding P-Cygni profile (step iv).  Moreover, an additional
blueshifted emission component can be easily identified in higher
resolution images (see Fig.~\ref{irlya}), as proposed in step v. Finally,
SBS~0335-052 shows a very broad, damped \lyal\ absorption line. This could
hint to a very young object (step i), or to a geometrical effect, if we
were observing from outside the ionization cone and through dense layers of
static neutral gas.

In this section we discuss in more detail the observed properties of
the different objects considered, and how they fit within the proposed 
scenario.  

\noindent
{\em SBS~0335-052:} SBS~0335-052 is a metal-deficient blue compact galaxy
which is hosting a very strong star-forming episode. Its morphology and
properties were described by \citet{thil97}. They
identified 6 super star clusters containing an equivalent of $\sim$4500
ionizing O7 stars. This galaxy is embedded in a large HI cloud with a mass
of $\sim 10^9 M_\odot$. The HI column density along the GHRS aperture
measured by these authors is very large, N(HI) = $7.0 \times 10^{21}$
cm$^{-2}$. The observational properties of SBS~0335-052 would be consistent
with step i: a very young starburst  starting to ionize the
surrounding medium while a significant amount of HI gas remains in front of
the massive stars and is still static with respect to the central HII
region. As a result, while a strong emission line spectrum is seen in the
optical, the spectral region around \lyal\ is completely absorbed, showing
a damped absorption profile, similar to the ones computed in
Fig.~\ref{voigt}.

\citet{klyal98} identified the OI and SiII absorption lines attributed to
the neutral gas at the same redshift than the central HII
region. \citet{thi97} postulated the presence of two additional OI and SiII
absorption systems blueshifted by 500 and 1500 \kms, which might indicate
that some layers of HI could be starting to be accelerated by the release
of mechanical energy from the starburst, as expected in step i (see
Fig.~\ref{model1}a).

Fig.~\ref{sbs1} shows the presence of a marginal excess emission in
the wings of the Voigt profile, suggesting an intrinsically very strong
\lyal\ emission line. We have fitted the intrinsic \lyal\ profile that
would produce such an excess, after convolution with the absorption
expected by the large amount of neutral gas, deriving a very strong
emission with $W(Ly\alpha) = 120$ \AA, consistent with a very young and
powerful starburst.

\noindent
{\em IZw18:} The GHRS high resolution spectrum of IZw18 was first discussed
by \citet{kunth94} and reanalyzed by \citet{klyal98}. It ressembles very
much the spectrum of SBS~0335-052, with a broad, damped \lyal\ absorption
profile corresponding to an HI column density $N(HI)$ $\sim 3.0 \times
10^{21}$ cm$^{-2}$. IZw18 is also a metal deficient blue compact galaxy
surrounded by a dense cloud of neutral hydrogen.  \citet{klyal98} showed
that the OI and SiII interstellar absorption lines were at the same
redshift than the central HII region, indicating that the neutral gas is
unperturbed and mostly static with respect to the starburst region.

As discussed above, no \lyal\ emission is detected along the STIS slit,
suggesting that a static neutral gas cloud completely covers the ionized
region with a large column density along the line of sight. This case seems
in principle to be similar to SBS~0335-052. Nevertheless, while the
starburst in this latter galaxy seems to be very young, there are hints for
a continued massive star formation activity in IZw18 over the last 10--15
Myr \citep{mhk99}, although at a rather weak rate. Moreover, large
supershells and ionized filaments have been identified in this galaxy by
\citet{martin96}. The observer might be looking in
this case through the large amounts of static neutral gas expected 
outside the conical HII region, which could be located along the
supershells  detected in the optical.

\noindent
{\em T1214-277:} The low resolution GHRS spectrum of T1214-277 has been
described in detail by \citet{thi97}. This very low metallicity galaxy
($\sim Z_\odot/23$) hosts a very young and massive starburst. As these
authors quote, this galaxy has a large \lyal\ equivalent width
($\sim$70 \AA).  \citet{ctm86} measured also a very large \hbeta\ equivalent
width ($W(H\beta) \sim 320$ \AA\ within a $2\arcsec \times 4\arcsec $
aperture).  No any  intrinsic extinction  was measured by 
these authors from the
Balmer lines ratios (but note that the Galactic extinction alone, with \ebv
= 0.06, destroys around 45\% of photons in the \lyal\ region). The high
equivalent widths and the absence of WR features \citep{pag92} indicate
that the starburst in T1214-277 has to be younger than 3 Myr, time at which
the first massive stars enter the WR phase at low metallicities \citep{mhk99}.

These observational results indicate that T1214-277 could be at the end of
step ii, as explained above: a conical HII region  would have already
developed, so that an
observer looking through it could detect the strong \lyal\ emission produced
by the central HII region without significant distortion.

In Fig.~\ref{t1214-la} we show a detail of the \lyal\ emission profile in
velocity scale. The profile is broadened, with 2 symmetric secondary peaks
located at $\sim \pm 300$ \kms\ from the central emission peak.  We
postulate that the starburst in this galaxy may be indeed already entering
step iii, so that recombination within the expanding shell
produces the additional \lyal\ components. Since only a fraction of the
shell is seen through the GHRS aperture, the observer detects only 2
additional peaks centered at $\pm$\vexp. The velocity derived for the shell
(300 \kms) is within the expected range (few hundred \kms) and is similar
to velocities measured in the other starburst galaxies with P-Cygni profiles.

\noindent
{\em Haro~2:} Haro~2 is a well studied, rather metal-rich blue compact
galaxy. As described by \citet{lequeux95} and \citet{klyal98} Haro~2 shows
a prominent \lyal\ emission line, with a clear P-Cygni profile. The
metallic OI and SiII interstellar absorption lines are detected at around
-200 \kms\ with respect to the central HII region, without any significant
absorption feature detected at the redshift of the HII region. Haro~2 would
be the prototype of our step iv, with a shell of neutral gas expanding at
around 200 \kms\ with respect to the central HII region, ploughing into a
low density, ionized halo without a screen of static neutral gas.

\begin{figure}
\epsscale{0.8}
\plotone{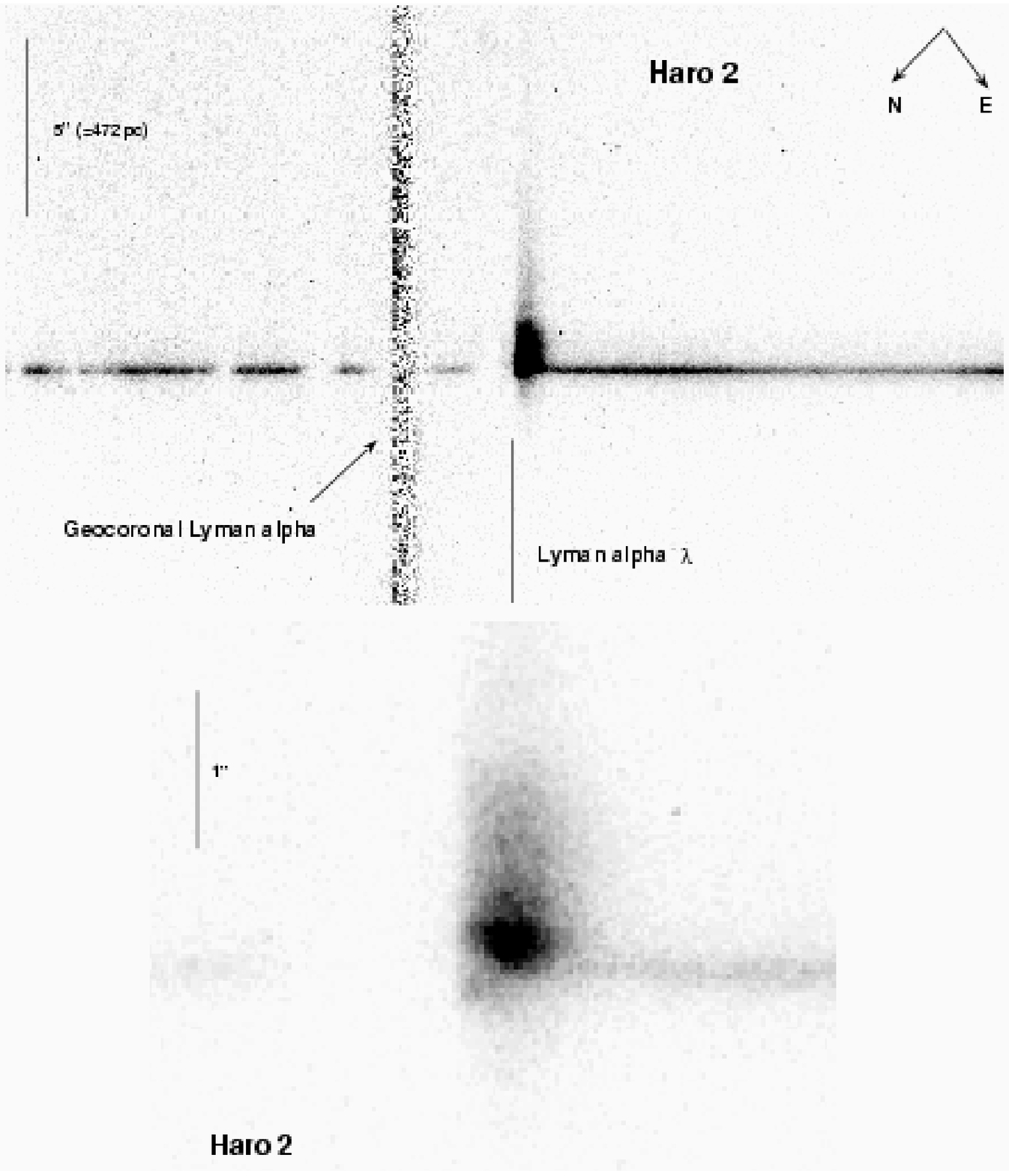}
\caption{UV spectral images of Haro~2. Top: we have marked the position of the
  geocoronal \lyal\ line residuals, after background subtraction. The
  vertical line marks the expected position of the \lyal\ wavelength at the 
  redshift derived from the HII region. The pixels have been rebinned for
  better display. We show the angular scale along the spatial axis, as
  well as the corresponding spatial scale. The N and E arrows indicate the
  orientation of the slit on the sky. Bottom: Detail of the \lyal\ region.  The
  image cuts have been selected to show the structure of the \lyal\
  emission. The displayed pixels correspond to physical detector
  pixels. The wavelength scale increases to the right on the X axis. }
\label{2dh2}
\end{figure}

\begin{figure}
\epsscale{0.7}
\plotone{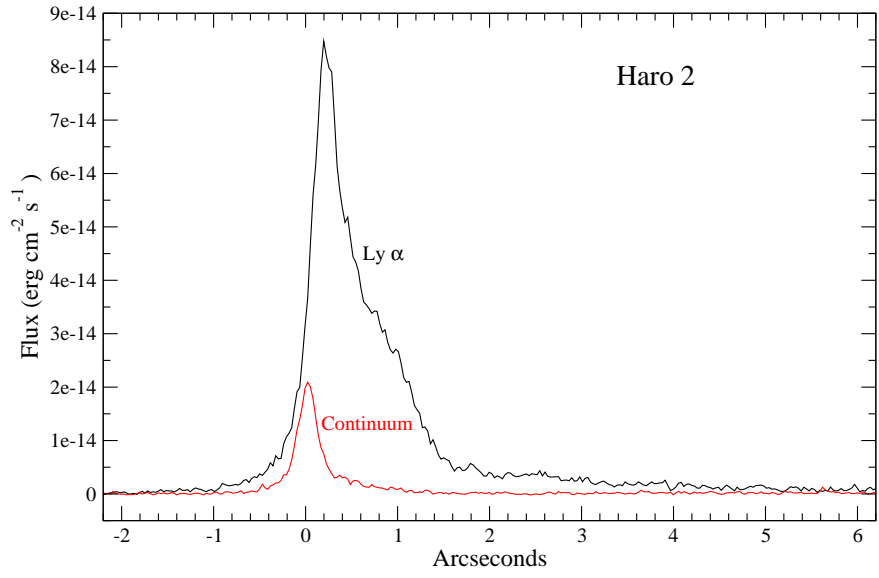}
\caption{Spatial profile of the \lyal\ line in Haro~2 on top of the UV 
  continuum profile. Note the \lyal\ emission extended to the SW (to the 
  right in the plot),   with
  almost no counterpart on the UV continuum.}
\label{h2lyalspat}
\end{figure}

\begin{figure}
\epsscale{0.7}
\plotone{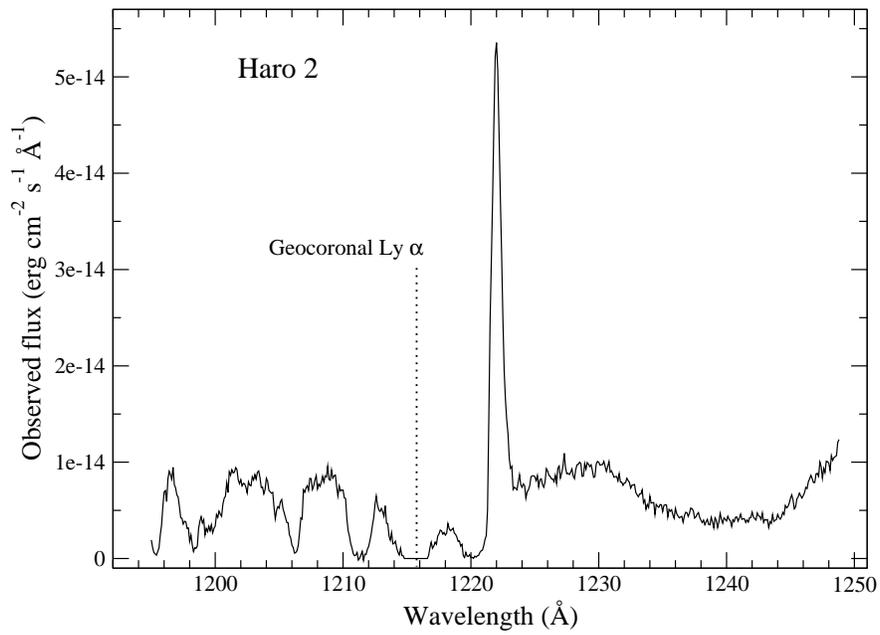}
\caption{Extracted spectrum of the central region of Haro~2 (corresponding
  to 0\farcs4). The
  position of the geocoronal \lyal\ line has been marked. Several
  interstellar absorption lines are detected bluewards of the Galactic
  \lyal\ line. Note the broad
  but weak apparent absorption centered around 1240 \AA, of unknown origin.}
\label{h2lya-1}
\end{figure}

\begin{figure}
\epsscale{0.7}
\plotone{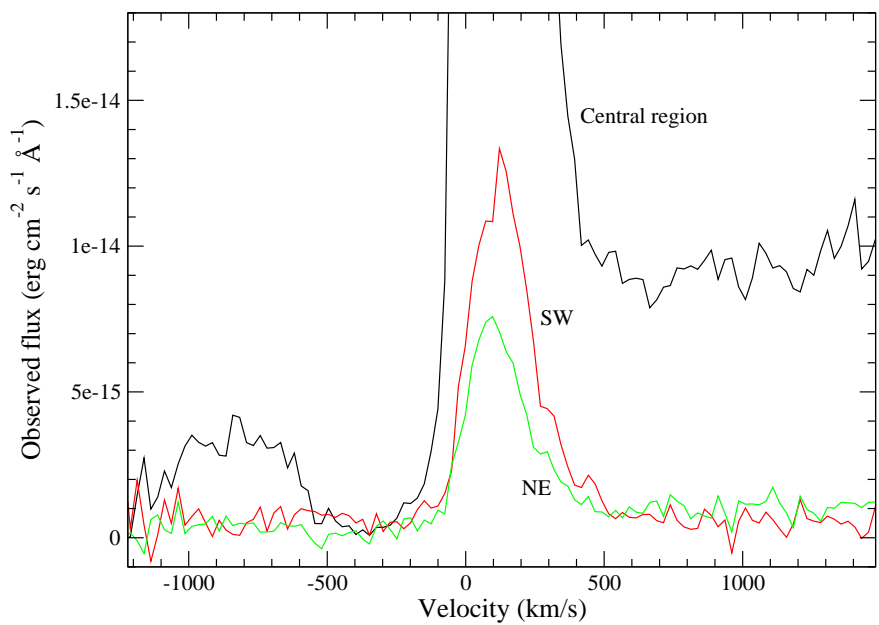}
\caption{\lyal\ emission line spectral profiles in different 
  regions of Haro~2. Strongest line: central region. 
  Intermediate profile: extended
  emission integrated over a region of 1\farcs5 centered 2\farcs3 SW of the
  nucleus. Weakest profile: extended emission over 0\farcs7 centered at
  0\farcs8 NE of the nucleus. Note that the profiles red wing vanishes
  always at around 500 km s$^{-1}$, independently of the strength of the
  line. Similarly, the blue wing goes to zero at around $-200$ km s$^{-1}$
  in the 3 regions.}
\label{h2lya}
\end{figure}

\begin{figure}
\epsscale{0.7}
\plotone{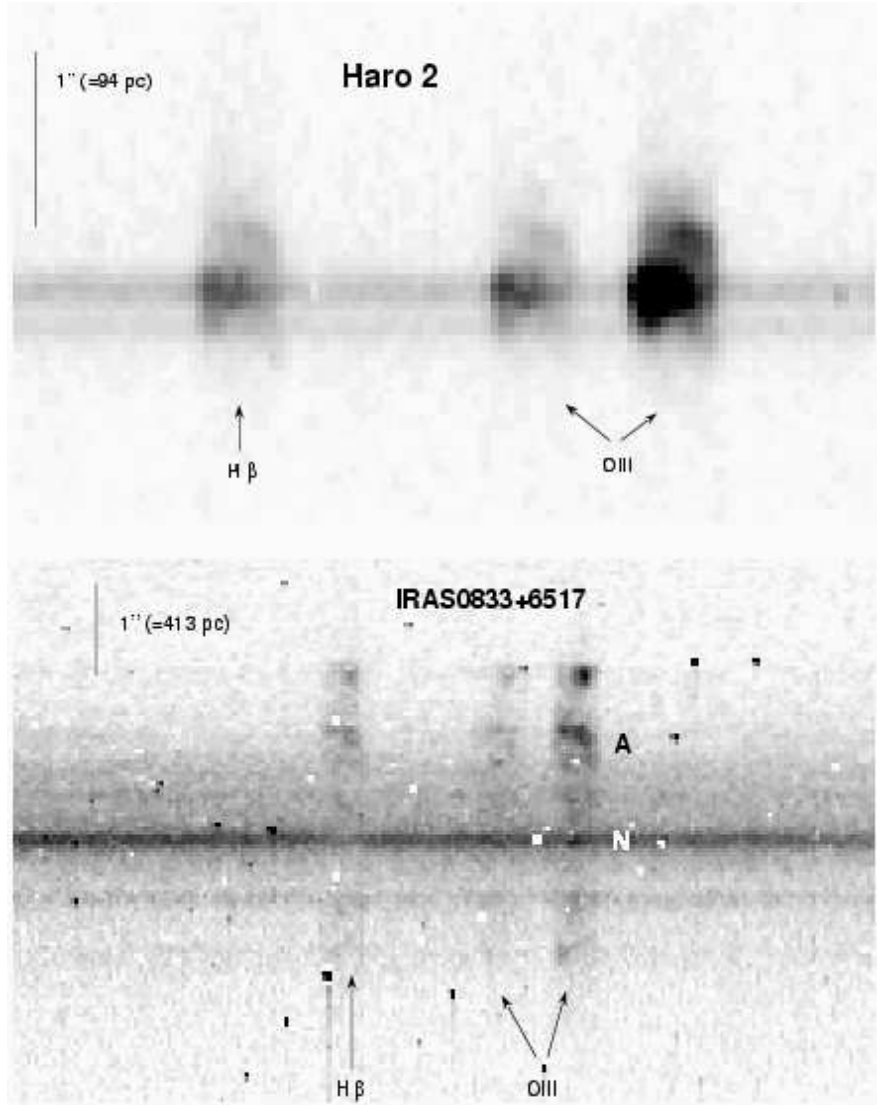}
\caption{Optical spectral images: Haro~2 (top) and IRAS~0833+6517 (bottom
  panel). The position of the \hbeta\ and
  OIII lines has been marked, as well as the angular and spatial scales
  along the Y axis. The pixels displayed 
  correspond to physical detector pixels. 
  Some velocity structure is apparent in the OIII lines on the Haro~2
  image. 
  The velocity shift
  between both peaks corresponds to around 700 km s$^{-1}$. Such velocity
  offsets have not been detected on the \lyal\ emission line. 
  In the bottom panel we have marked with 'A' and 'N'
  the regions for which we have extracted the spectra shown in
  Fig.~\ref{irhb}. Dark and white pixels are the residuals of regions
  affected by cosmic ray hits. }
\label{h2-o2d}
\end{figure}

\begin{figure}
\epsscale{0.7}
\plotone{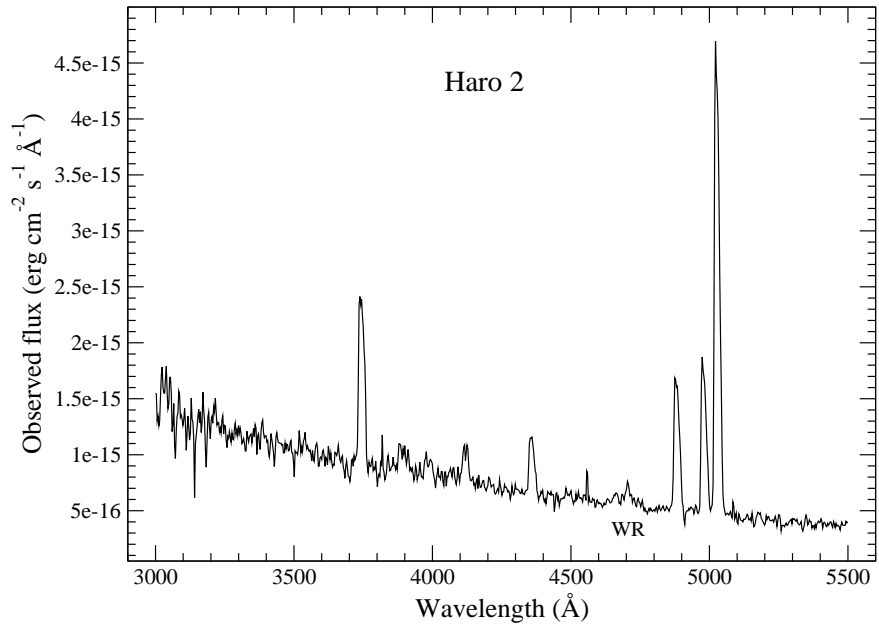}
\caption{Optical spectrum of Haro~2, integrated over an aperture of
  0\farcs3. Note the Wolf-Rayet feature at around 4686 \AA. }
\label{h2hb}
\end{figure}

From the STIS long slit spectral images (Fig.~\ref{2dh2})
we can identify all the features expected in step iv:

\begin{itemize}

\item The emission decreases rapidly bluewards of the velocity 
 at which the
emission peak originated from the central HII region.
The flux goes to zero at basically the same
velocity all along the slit. As discussed in Sect. 2, this is consistent
with an expanding shell of neutral gas which has to be much more extended
than the 8\arcsec\ over which the absorption is detected in order to behave
as a plane-parallel slab. From the analysis of \hal\ long-slit spectroscopy
\citet{legrand97} found evidence for the presence of an ionized expanding
shell over a region around 25\arcsec\ in diameter (nearly 2.4 kpc). The
lateral edges of the shell appeared at the same systemic velocity than the
central HII region, decoupled, as expected, from the disk rotation. The
\hal\ line is also clearly broadened at the edges, in good agreement with
what would be expected from an expanding shell.

As explained above, step iv foresees that as the starburst evolves, the
trapping of the ionization front will yield the formation of a neutral
layer at the external side of the expanding shell, with no static neutral
gas inside the shell. This would produce the kind of profile we detect on
the central region of Haro~2 (see Figs.~\ref{h2lya-1} and \ref{prof1}).

\item At the same time, recombination on the internal layers of the shell 
produces
additional emission components at $\pm$\vexp, in this case at $\pm$200
\kms. While the blueshifted \lyal\ emission component produced in the
internal layers will be destroyed by resonant scattering on the external
neutral layers (both expanding at the same velocity), this blueshifted
emission should be detectable in \hal. \citet{legrand97} concluded indeed
that the \hal\ emission profile (when looking straight into the starburst
region) showed  evidence for two additional weak components at $\pm$200 \kms. 

These results indicate that a section of the shell receding at the same
velocity should also be present. The internal part of the receding shell
would produce \lyal\ photons, as it generates \hal\ ones. And the neutral
layers of the receding shell would backscatter in addition some \lyal\
photons produced at the central HII region. Since the receding neutral gas
lies at -\vexp\ with respect to the central HII gas it scatters the
blue wing of the intrinsic \lyal\ emission profile. A key point is that
this receding shell lies at $+2\times$\vexp\ with respect to the
approaching shell, so that the \lyal\ photons emitted or backscattered 
  there
 traverse unaffected the approaching neutral layers and reach the
observer redshifted by $+$\vexp. As a result, we expect an extended, broad
and weak component of \lyal\ photons centered at around +200 \kms\ in the
case of Haro~2 (see the example in Fig.~\ref{bc1}). 
 
This is exactly what we detect in our spectral images (Fig.~\ref{2dh2}). A
weak extended emission is detected at both sides of the central HII region
on the slit, but especially to the upper side. It results from the \lyal\
photons originating from or backscattered by the receding shell. We show in
Fig.~\ref{h2lya} the profiles of this emission at several arcseconds from
the central HII region. The profiles show:

\begin{itemize}

\item In both parts of the slit, SW and NE of the central HII region,
the blue edge of the emission profile lies at the same velocity: the
approaching neutral shell is destroying all photons blueshifted  with
respect to the velocity of the central HII region.

\item Broad extentions up to around $+500$ \kms. Profiles are
more symmetric than the central \lyal\ emission line and their peaks are
clearly redwards of the \lyal\ peak. Both profiles show quite extended red
wings at basically the same velocity. These wings
originate by recombination and backscattering in the ionized and neutral
layers of the receding shell, respectively.  Since the scattering process
affects photons within a high range of velocities, as shown in
Fig.~\ref{h2lya}, the backscattered component should cover basically the
same velocity range measured on the absorption profile. We consider indeed
that the good agreement between the terminal velocity of the P-Cygni
profile (around $-$500 \kms) and the terminal velocity of the extended
\lyal\ emission profile (around $+$500\kms) supports our interpretation.

\end{itemize}
\end{itemize}

\begin{figure}
\epsscale{0.8}
\plotone{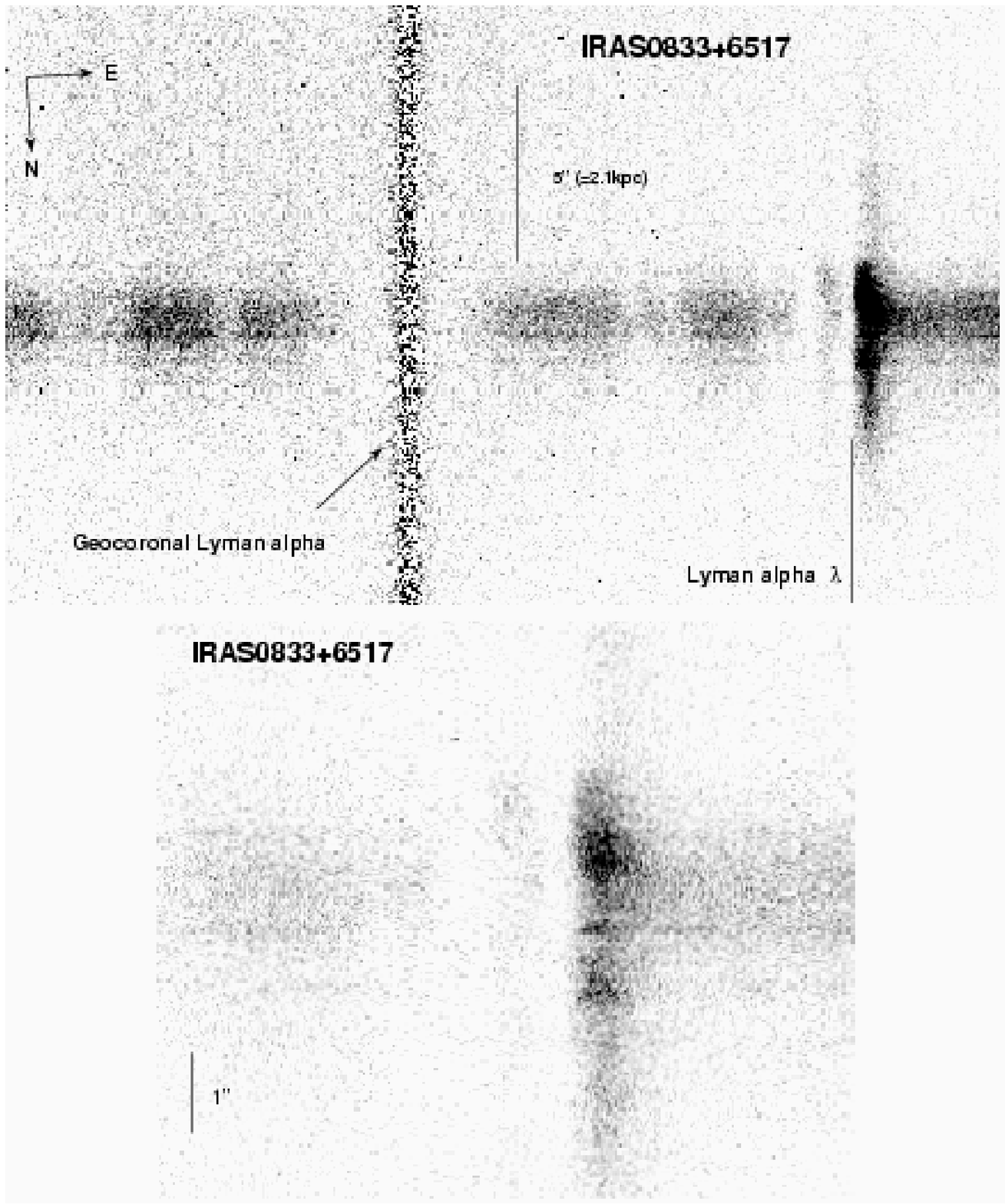}
\caption{
UV spectral images of IRAS~0833+6517. Top: we have marked the position of the
  geocoronal \lyal\ line residuals, after background subtraction. The
  vertical line marks the expected position of the \lyal\ wavelength at the 
  redshift derived from the HII region. The pixels have been rebinned for
  better display. We show the angular scale along the spatial axis, as
  well as the corresponding spatial scale. The N and E arrows indicate the
  orientation of the slit on the sky. Bottom: Detail of the \lyal\ region.  The
  image cuts have been selected to show the structure of the \lyal\
  emission. The displayed pixels correspond to physical detector
  pixels. The wavelength scale increases to the right on the X axis. 
  The extracted spectrum is shown in Fig.~\ref{irlya}.}
\label{2dir}
\end{figure}

\begin{figure}
\epsscale{0.7}
\plotone{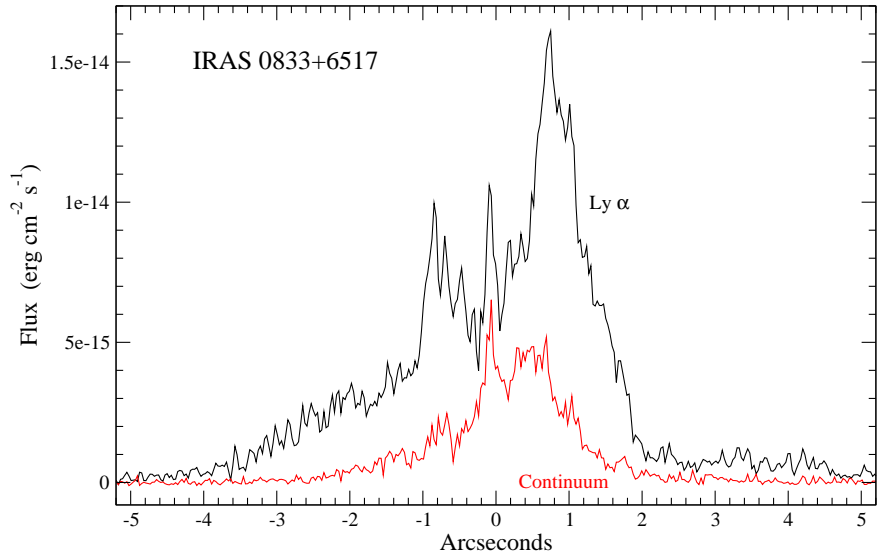}
\caption{Spatial profile of the \lyal\ line in IRAS~0833+6517 on top of the UV 
  continuum profile. The regions with the strongest \lyal\ emission are not
  correlated with the location of the strongest UV continuum sources.} 
\label{irlyalspat}
\end{figure}

\begin{figure}
\epsscale{0.7}
\plotone{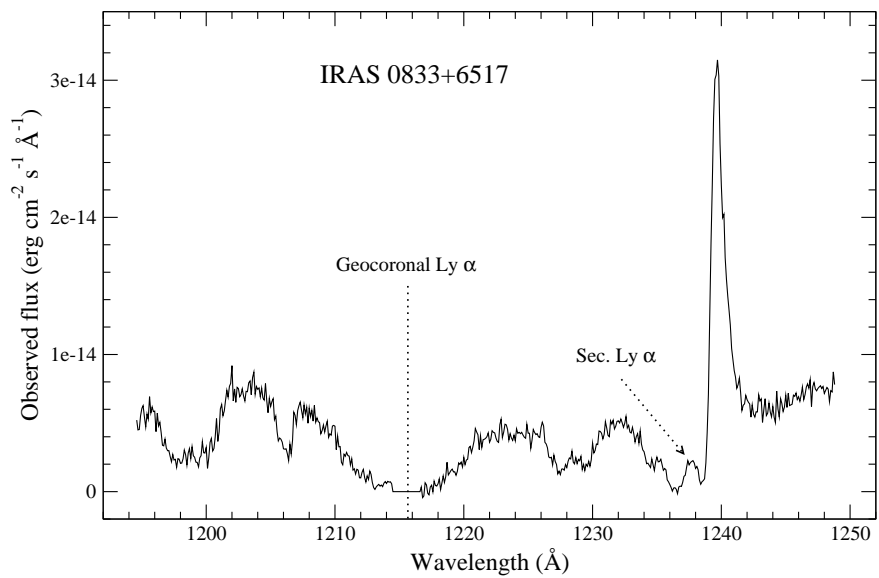}
\caption{Extracted spectrum of the central region of IRAS~0833+6517
  (corresponding to 2\farcs0). The position of the geocoronal \lyal\ 
  line has been marked,
  in the center of the broad Galactic absorption profile. We have also
  marked the position of the secondary emission peak, whose spatial
  distribution can be appreciated in Fig.~\ref{2dir}. Note the splitting of
  the SiIII $\lambda$1206 absorption line, with peaks at 1227 and 1229
  \AA. Additional broad interstellar absorption lines are detected
  bluewards of the Galactic \lyal\ line. }
\label{irlya}
\end{figure}

\begin{figure}
\epsscale{0.7}
\plotone{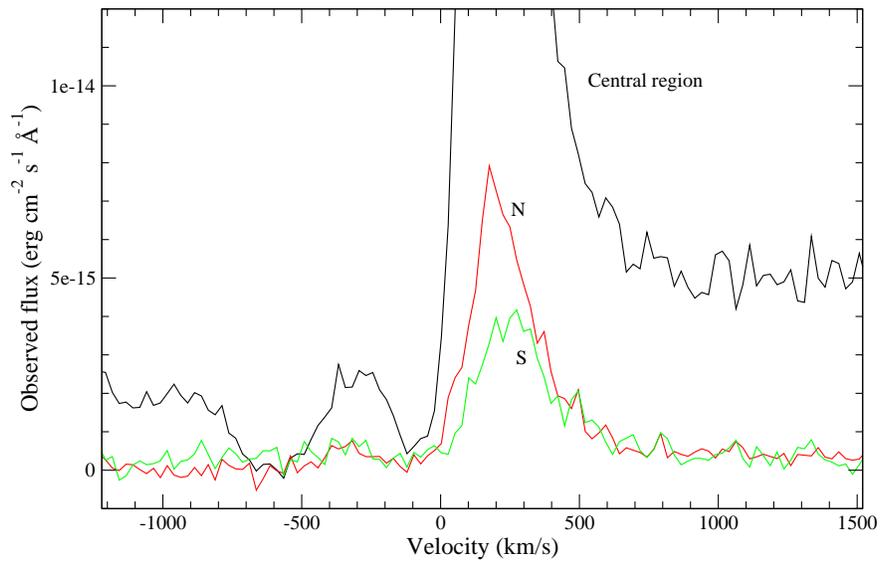}
\caption{\lyal\ emission line spectral profiles in different IRAS~0833+6517
  regions. Strongest line: central region. Intermediate profile: extended
  emission integrated over a region of 2\farcs3 centered 3\farcs0 N of the
  nucleus. Weakest profile: extended emission over 4\farcs1 centered at
  3\farcs0 S of the nucleus. Note again that the profiles extend over
  basically the same velocity range independently of their strength. The
  secondary emission profile, centered at $-300$ \kms\ appears only in the
  central region. }
\label{irlya-2}
\end{figure}

\begin{figure}
\epsscale{0.7}
\plotone{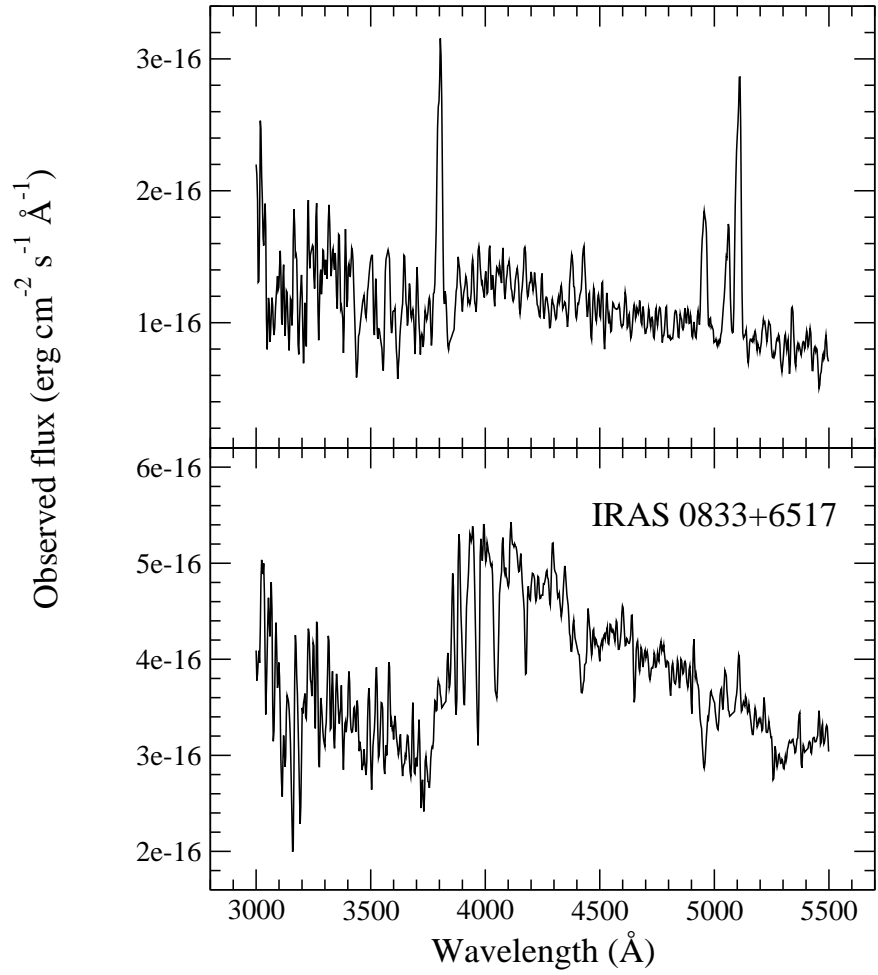}
\caption{Extracted optical spectra of IRAS~0833+6517 for the regions marked
  in Fig.~\ref{h2-o2d}: 'A' region in the top panel, and 'N' region  
  in the bottom one. }
\label{irhb}
\end{figure}

\begin{figure}
\epsscale{1.0}
\plotone{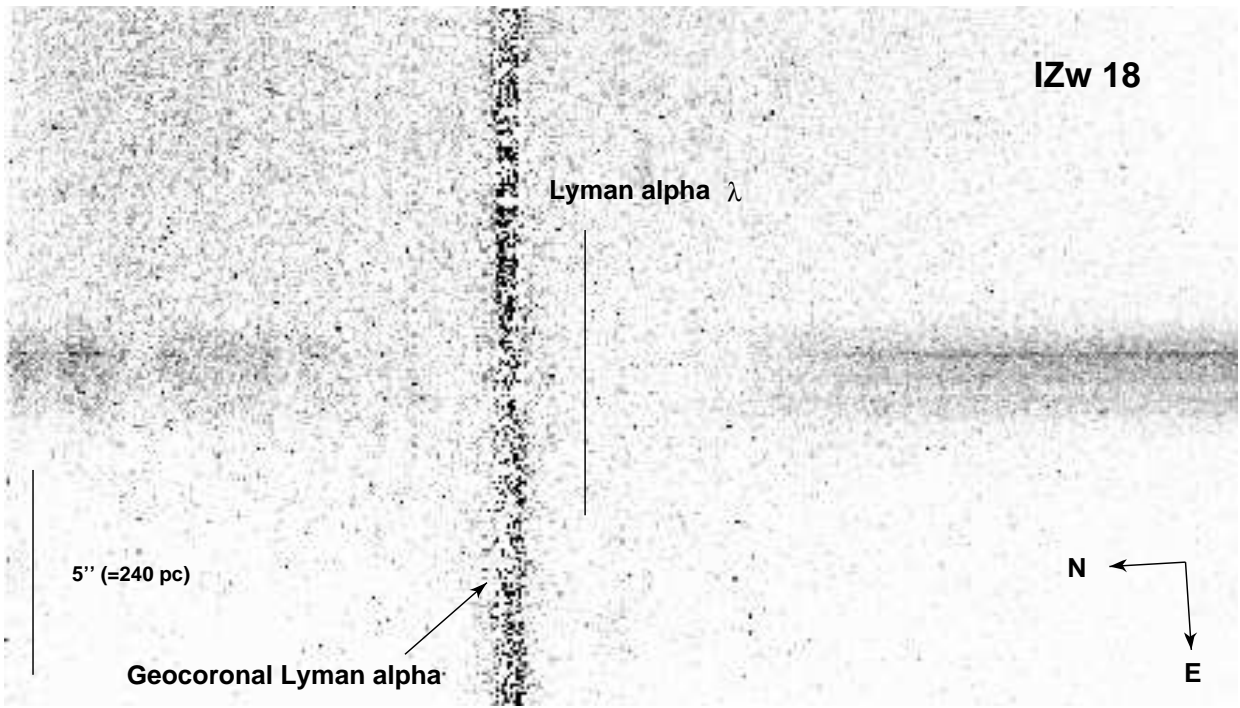}
\caption{UV spectral image of IZw18. Symbols as in Fig.~\ref{2dh2}. Note the
  broad damped absorption profile, blended with the Galactic
  absorption. Note also the lack of \lyal\ photons in emission along the
  slit.  }
\label{2diz}
\end{figure}

\begin{figure}
\epsscale{0.7}
\plotone{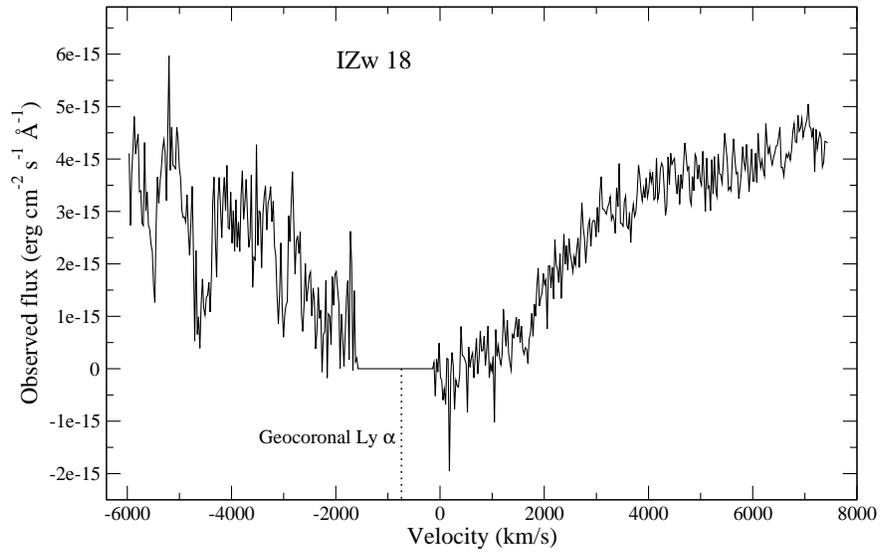}
\caption{Extracted spectrum of the central region of IZw18 (corresponding
  to 2\farcs0). The
  abscissa axis is given in velocity scale to show the width of the absorption
  profile. Additional interstellar absorption lines are detected bluewards
  of \lyal. }
\label{izlya}
\end{figure}

\begin{figure}
\epsscale{0.7}
\plotone{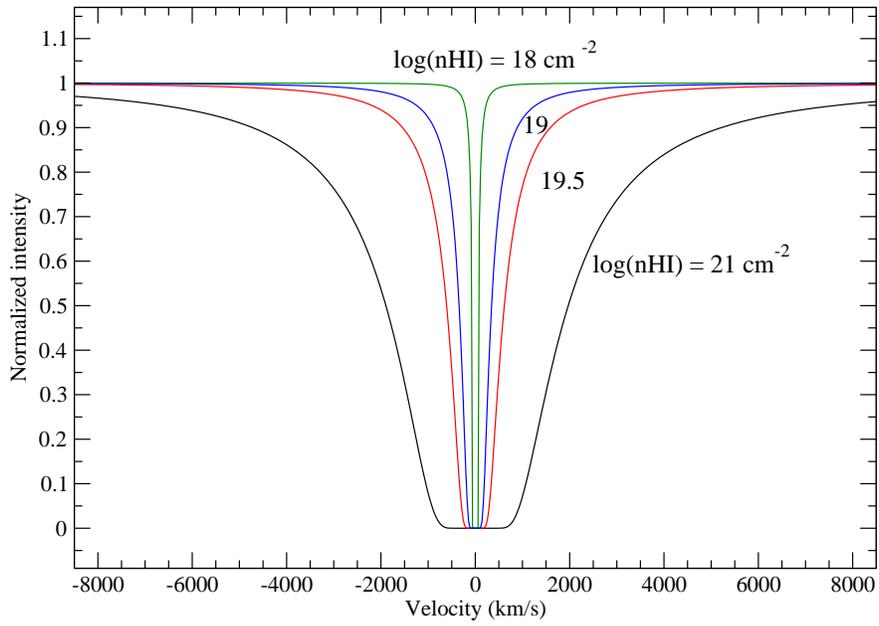}
\caption{Voigtian absorption profiles computed for neutral hydrogen densities
  log(nHI) = 18, 19.5, 20 and 21 cm$^{-2}$ (from weaker to stronger
  absorption). We have assumed $b$ = 20 \kms.}
\label{voigt}
\end{figure}

\begin{figure}
\epsscale{0.7}
\plotone{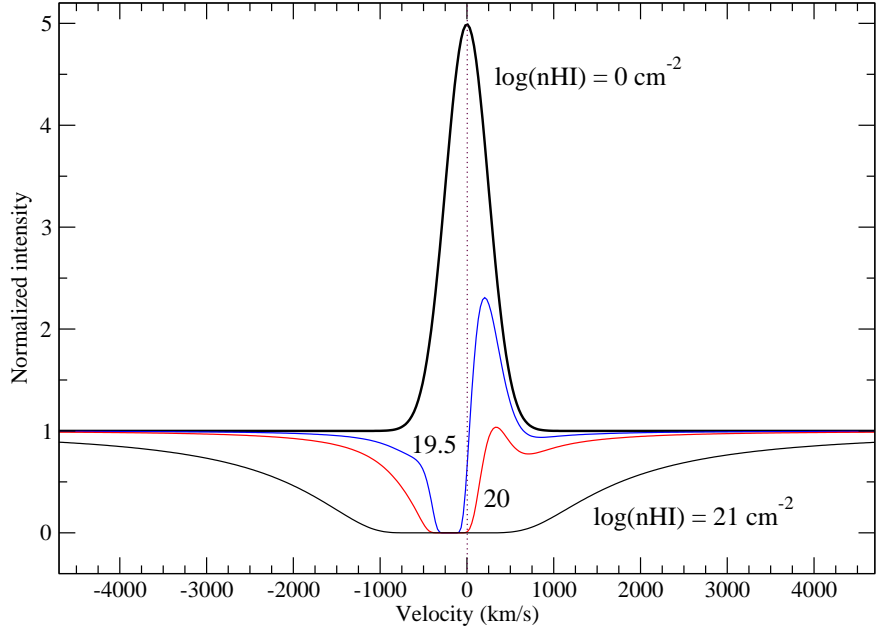}
\caption{Expected \lyal\  profiles. We have assumed an intrinsic \lyal\ 
  emission line originating in the central HII region, 
  at systemic velocity
  (V = 0 \kms). The plot shows the intrinsic emission profile (thick
  line, log(nHI) = 0 \kms) 
  and the resulting profiles assuming a slab of neutral hydrogen moving at
  \vexp\ = $-200$ \kms, with column densities log(nHI) = 19.5, 20 and 21
  cm$^{-2}$. }
\label{prof1}
\end{figure}

\noindent
{\em IRAS~0833+6517:} As discussed in Sect.~2, \lyal\ emission
in IRAS~0833+6517 has a more
complex structure than in Haro~2 (see Fig.~\ref{2dir}). In addition a 
secondary \lyal\ emission is conspicuous on top of the damped absorption.
Nevertheless, the overall properties of the \lyal\ emission are very
similar between both galaxies:

\begin{itemize}

\item The flux vanishes  bluewards of  the same wavelength along the slit 
(hence the same velocity)  whenever \lyal\ emission is detected (10\arcsec, 4.1
  kpc). The extension of the neutral gas shell has  to be very large 
in this  galaxy, reaching a diameter close to 10 kpc. 

\item S and N of the central region where \lyal\ is in emission, 
we detect again an extended broad and weak
  component redshifted with respect to the
  central HII region by around 300 \kms. The terminal velocity of the
  \lyal\ profile all along the slit is at around $+$700 \kms. As in Haro~2,
  we associate
  this component to the emission originated at the receding shell, both by
  the ionized inner layers and by backscatter in 
the neutral external layers. 

\end{itemize}

We therefore argue that the superbubble phase in IRAS~0833+6517 corresponds
to step iv in our model. The presence of an additional emission component
on top of the absorption profile blueshifted by around $-$300 \kms\ with
respect to the central HII region, supports our interpretation that an
expanding shell is moving towards the observer at this velocity. As
explained in step v, the leading shock in this galaxy should have become
already radiative, undergoing recombination and originating the secondary
\lyal\ emission line we detect at the velocity of the shell. The fact that
this secondary emission does not appear affected by neutral gas scattering
indicates that the HI column density in front of the expanding shell
(either static or also expanding) has to be rather small, as expected in
our model. The smaller spatial extension of this secondary emission
indicates that only a fraction of the whole surface of the leading shock is
undergoing recombination. Conditions given in step v strongly depend on the
density of the halo, which is most likely not uniform, explaining the
patchiness of this emission.

We have overplotted in Fig.~\ref{iras08l-h} our high resolution STIS
spectrum of IRAS~0833+6517 over the lower resolution GHRS one presented by
\citet{rosa98}. This comparison exemplifies how the analysis of P-Cygni
profiles at low resolution may yield misleading results. The asymmetry of
the intrinsic profile translates into an artificial redshift of the
emission line peak when observed at low resolution (in addition to the
effect discussed in Sect.~3, Fig.~\ref{prof2}). This effect might be
especially severe when analyzing galaxies at high redshift with medium or
low resolution spectra. Furthermore, the lower resolution does not allow to
identify the damped, black absorption plateau (which in this case is
partially filled by the secondary emission peak), preventing an accurate
derivation of the intervening HI column density.

We want to remark that our model provides a simplified scenario for the
formation of \lyal\ P-Cygni profiles in starburst regions. The actual
geometry and kinematical structure of some objects, like IRAS~0833+6517,
might be significantly more complicated, although the overall view should
still be valid. Let us highlight some observational properties that may be
relevant:

\begin{itemize}

\item \citet{klyal98} were not able to detect the interstellar OI
$\lambda$1302.2 and SiII $\lambda$1304.4 lines in their high resolution
GHRS spectrum. The analysis of the lower resolution data from
\citet{rosa98} shows that both lines are blended within a very broad
absorption profile, which in fact was extended over most of the
\citet{klyal98} GHRS wavelength range. \citet{rosa98} report also very
broad profiles on other interstellar lines, with FWHM up to around 1000
\kms. They interpret these broad absorption profiles as evidence of
large-scale motions of the interstellar gas around the starburst region.

\item The SiIII $\lambda$1206 absorption line detected bluewards of 1230
\AA\ is split into two components (see Figs.~\ref{2dir} and
\ref{irlya}). While the weakest component is centered at the redshift of
the central HII region, the strongest one appears blueshifted by around
$-$470 \kms, which is higher than the expansion velocity attributed to the
shell. \citet{rosa98} also found from lower resolution GHRS spectra that
the interstellar SiII$\lambda$1260 and CII $\lambda$1335 were blueshifted
by 450 -- 520 \kms\ with respect to the central HII region. We want to
stress that no \lyal\ emission at around $-$500 \kms\ can be
detected. Therefore the high velocity gas has to be contained within the
expanding shell.

\end{itemize} 

We conclude that while the bulk of the neutral gas shell seems to be
expanding at around 300 \kms, some gas might be moving within the shell in
a rather chaotic way at significantly higher velocities.

\noindent
{\em Other galaxies:} The model we have proposed could also explain the
observational properties of the \lyal\ line in other starbursts studied up
to now. IIZw70 and Mrk36, discussed by \citet{klyal98}, and Tol~65,
presented by \citet{thi97}, would be similar to IZw18 and SBS~0335-052:
they show a damped broad absorption profile, and their interstellar
absorption lines appear at the redshift of the central HII region. The HI
gas is therefore static along the line of sight with respect to the HII
region, either because their starburst episodes have not been able to
accelerate the gas (being very young or weakly energetic compared to the
amount of neutral gas surrounding them), or because we are observing them
through the densest part of the HI disk. In this latter case \lyal\
emission could possibly leak from other regions (for instance if an
ionizing cone is allowing the photons to escape) and they might be
scattered to our line of sight.

We are presently performing a survey with HST--ACS looking for regions in
this kind of galaxies where \lyal\ photons could be leaking. Preliminar
results  appear in \citet{acs1}. 

Other galaxies studied by \citet{klyal98}, like ESO~400-G043 and
ESO~350-IG038, showing P-Cygni \lyal\ absorption and blueshifted
interstellar lines, would also be in the state described by our step iv,
which seems to be the most frequent one among the galaxies showing \lyal\
emission.  

We want finally to stress that the expanding layer causing the typical
P-Cygni \lyal\ line profile in starburst galaxies is extended and can span
several kpc across a galaxy. As discussed above, in some objects this can
be traced across several hundreds of pc away from the main burst of recent
star formation. In the case of IRAS 0833 this spans more than 2.2 kpc
across the galaxy.  Our observations thus prove that this is not a well
localized phenomenon, but rather a large scale one affecting the whole
central region of these galaxies.  Nevertheless, to explain the properties
of this emission (lack of structure and clumpiness, velocity dispersion and
intensity) the low density medium or halo of the host galaxy, into which
the shell propagates, has to be very extended so as to allow for the
existence of a continuous shell with similar properties (column density,
velocity, etc...) over dimensions $\geq$ 1 kpc. If this were not the case
and the expansion had reached the galaxy edge, upon blowout the shell would
have become Rayleigh-Taylor unstable and would have rapidly broken while
destroying the absorbing layer.  Thus, although the \lyal\ asymmetric line
profiles are indicative of strong outflows, these do not certify that
supergalactic winds, venting the metals from the newly formed starbursts
into the inter-galactic medium, have already developed.  In fact if the
expanding shell could reach the edge of its host galaxy, becoming
Rayleigh-Taylor unstable and fragmenting, the \lyal\ line profile to be
observed would be the almost unattenuated line produced by the central HII
region, without any absorption by neutral gas but not the kind of profiles
generally observed.

\section{Implications for galaxies at high redshift}

It is expected that all $L^*$ galaxies have undergone a phase of rapid star
formation at a certain stage of their evolution. During this phase they
should have become very powerful producers of ionizing photons. Very strong
emission lines are therefore expected from these objects, and with rather
large equivalent widths. For redshifts above $z \sim 3$, the \lyal\  emission
line would be observable in the optical range, making it therefore {\em
a priori}  an  optimal tracer of star formation. Nevertheless, the first
searches did not detect \lyal\  emitters. It has been only recently that
\lyal\  emission from galaxies at redshifts above $z \sim 3$ has been
detected, but at average luminosities weaker by 2 orders of magnitude than
expected (see \citet{rho00} for a complete set of references concerning
\lyal\  emission at high redshift).

Our analysis of \lyal\  emission line properties in nearby starburst galaxies
has shown that there are several issues affecting the shape and intensity
of the line, so that only in a small fraction of the cases the full
intrinsic emission would be detectable. In this section we will extrapolate
our results to high redshift galaxies, aiming to estimate the expected
properties of \lyal\  emission in these objects. 

Star-forming episodes detected in galaxies at redshift $z \sim 3$ seem to
be significantly stronger than those detected in compact galaxies in the
nearby Universe (see \citet{shap01}, \citet{shap03} and references
therein). Nevertheless, the properties of the \lyal\ emission profiles are
very similar to those observed in local starbursts.  
 Lyman Break Galaxies exhibit a broad distribution of \lyal\ strengths
 and profile-types, ranging from damped absorptions to pure emission,
 including also P-Cygni-like absorption and emission combinations
 \citep{shap03}.  
Asymmetric P-Cygni
profiles have also been detected by \citet{frye02},
\citet{pet00,pet01}, among other authors (see also \citet{heck01} and
\citet{rho00}).

\begin{figure}
\epsscale{0.7}
\plotone{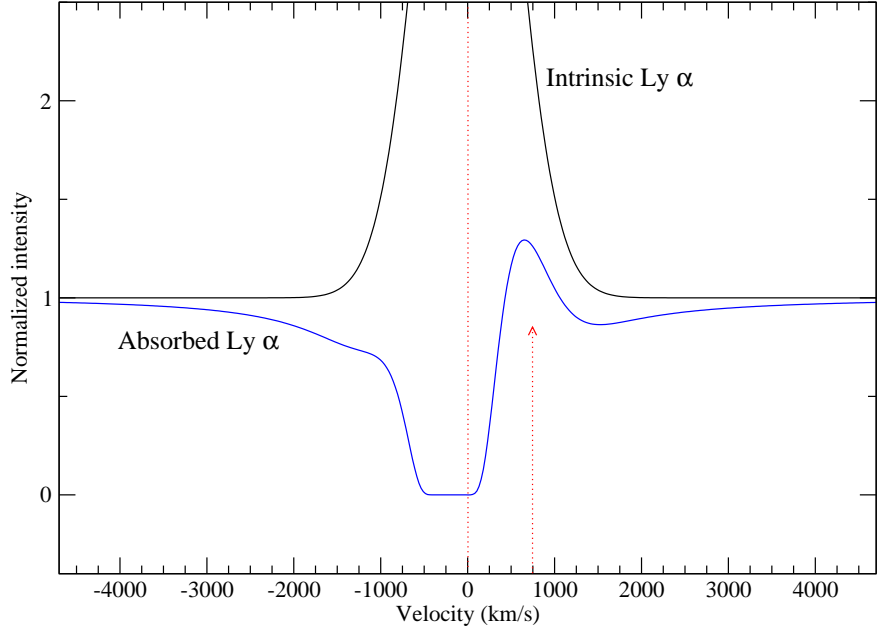}
\caption{Detail of a resulting \lyal\  profile showing that under certain
  circumstances the centroid of the observed \lyal\  emission line can appear
  redshifted by several hundred \kms. This redshift, nevertheless, is
  artificially originated by the absorption of the blue part of the profile,
  and should not be considered as a tracer of ionized gas
  outflows. Moreover, it is important to note that the width of the
  resulting line is much smaller than that of the intrinsic emission line. This
  example has been computed for log(nHI) = 20.3 cm$^{-2}$ and \vexp\ =
  $-300$ \kms. } 
\label{prof2}
\end{figure}

\citet{pet01} observing Lyman Break galaxies found
velocity offsets of the interstellar absorption lines with respect to the
velocity of the HII region, assumed to be at the systemic velocity of the
galaxy. They  found  blue velocity
offsets between approximately $-200$ and $-400$ \kms, 
in three-quarters of the objects,  with a median value
of $-300$ \kms. 
More recently, \citet{shap03} have obtained a composite spectrum by
  combining the data 811 individual  Lyman Break galaxies. This spectrum
  reveals that the strong low-ionization interstellar features appear
  blueshifted with respect to the systemic velocity by an average of $-150
  \pm 60$ \kms. Moreover, the mean velocity of the interstellar SiIV doublet 
  is close to $-180$ \kms,
  indeed very similar to the value measured for low ionization lines.
As discussed above, the presence of blueshifted
interstellar lines hints directly to the presence of neutral shells
expanding at few hundred \kms, as concluded by \citet{pet01} and
\citet{shap03}.  On the other hand the simultaneous detection of SiIV 
at roughly the same  velocities suggests that the internal layers of the 
expanding shells are  probably ionized, as discussed above. 
This range
of velocities is consistent with the predictions for a wind-blown shell, as
we have seen. On the other hand, these authors found the \lyal\  emission
lines to be always redshifted with respect to the HII regions. The offsets
measured for \lyal\  in the galaxies showing blueshifted interstellar
absorptions span a range between around $+$200 and around $+$600 \kms. 
 The composite spectrum of \citet{shap03} gives  a mean redshift
 of $+360$ \kms\ for
 the \lyal\ peak with respect to the systemic
  velocity. 
As explained above this is exactly what one  expects if the central
starbursts were driving a neutral (or partially ionized) expanding shell
(see Fig.~\ref{prof2}). 

\citet{pet00} show a high resolution, high signal to noise ratio spectrum
of MS 1512--cB58, a galaxy at $z = 2.73$. They found a weak and asymmetric
\lyal\  emission line, redshifted by around $+230$ \kms\ with respect to the
HII region, on top of a damped absorption with $N$(HI)$= 7.5 \times
10^{20}$ \cm, blueshifted by $-390$ \kms\ (in agreement with the
interstellar lines). At this rather high HI column densities the absorption
profile is very broad, allowing only a small fraction of the intrinsic
\lyal\  emission line photons to escape. A careful analysis of their Fig.~4
shows the presence of a broad red wing, extending up to $\sim 800$ \kms
(velocity measured with respect to the HII region). We postulate that this
red wing component might be the emission originated at the receding part of
the shell, either by recombination or by backscattering of \lyal\  photons
from the central HII region, as discussed above. We expect this component
to be broader the higher the HI column density, since backscatter would
affect photons at increasingly high velocities. The presence of the
broad red wing provides additional support to the existence of a rather
symmetrical expanding shell being energized by the central starburst. 

Asymmetric \lyal\  emission profiles have also been detected by
\citet{frye02} on a sample of eight lensed galaxies at high redshift, $3.7
< z < 5.2$. Interstellar absorption lines have also been detected in some
of these galaxies, but unfortunately the redshift of the central HII region
is not available. They found an offset between the redshift
of the interstellar lines and that of the \lyal\  emission centroid in the
range 300--800 \kms. If we assume that the interstellar lines are tracing a
neutral shell expanding at around 200--300 \kms\ with respect to the central
HII region, the \lyal\  lines detected would be redshifted by up to around
400--500 \kms\ at most, in agreement with the predictions and with other
galaxies. Note that \citet{frye02} always quote the
offset between \lyal\  and the interstellar lines as to be the local 
redshift of
the \lyal\  profile, something which is misleading since
 the shift with respect to the HII region has to be
necessarily smaller. Finally, most of the \lyal\  profiles plotted by
\citet{frye02} in their Fig.~13 show a weak broadening of their red wing,
in agreement with the model discussed in this work. 

 Another important result obtained by \citet{shap03} from their
analysis of Lyman Break Galaxies is that the \lyal\ emission strength
increases as the kinematic offset between the \lyal\ emission peak and the
low-ionization interstellar absorption lines decreases. In addition, the
\lyal\ equivalent width in emission increases as the interstellar lines
become weaker. We interpret these effects as an evidence that the lower the
neutral gas column density, the smaller the relative redshift of the
\lyal\ emission peak and, correspondingly, the stronger the emission. This
is in very good agreement with our above discussion (see for example
Fig~\ref{prof2}).

We  therefore conclude that the properties of the \lyal\  emission profiles
are very similar in local starbursts and in high redshift galaxies.
Nevertheless, there are some differences between both kinds of objects,
which have to be taken into account for a correct interpretation of the
data:

\begin{itemize} 

\item Since the star-forming episodes observed at high redshift are very
  powerful ones, we expect intrinsically very strong \lyal\  emission lines,
  with emission extending  over several
  hundreds \kms\ at the base of the line.
 Even accounting for HI column densities around
  10$^{20}$ \cm,  photons at the red wing of the emission profile will
  be able to escape. Therefore, most of these high redshift galaxies
  experiencing strong starbursts should show \lyal\  emission, albeit reduced
  by large factors from their intrinsic luminosity. Only galaxies with HI
  column densities above $\sim 10^{21}$ \cm\ will  show  a damped \lyal\ 
absorption.   

\item For the same reasons we expect the observed \lyal\  emission lines to
appear redshifted with respect to the systemic velocity of the galaxies by
a broad range of velocities, up to several hundred \kms. This effect
should be smaller in local starbursts, since the \lyal\  emission lines would
be intrinsically weaker (and therefore narrower at their base).  If no
other line is available to measure the redshift, one should bear in mind
that the derived value is only an upper limit of the galaxy redshift, and
that the real one might be indeed smaller by several hundred \kms.
  
\item The spectra presented by \citet{frye02} show a feature which only
appears for galaxies at high redshift: the \lyal\  forest, that significantly
lowers the average continuum bluewards of \lyal. As a result, the blue
wing of the absorption produced by the local expanding shell can not be
identified in low resolution data. 

On the other hand, the presence of asymmetric \lyal\ profiles together with
a discontinuity in the continuum (the average continuum redwards of \lyal\
being stronger than bluewards of the line) should allow to identify 
\lyal\ in surveys aiming to estimate redshifts of galaxies at high redshift.

\item Stellar \lyal\  absorption lines might be important at low resolution,
  artificially decreasing the observed intensity of the \lyal\  emission
  line. At high resolution the stellar profiles can be detected and taken
  into account when measuring the line flux, but at low resolution both
  profiles might be blended  and the emission line
  substantially hidden (see \citet{valls93}).

\end{itemize} 

It is clear from the above discussion that using the \lyal\ emission line
as a tracer of star formation episodes might lead to severe errors. We
remark that both the measured line intensity, and thus the star formation
rate derived from it, as well as its velocity, have to be considered as
respectively lower and upper limits only. The problem might be especially
significant in the case of star formation rates derived from \lyal\
luminosities, which might be underestimated by more than an order of
magnitude.

From the above discussion we see that P-Cygni \lyal\ profiles are predicted
only when the supperbubble has entered the phase in which the ionisation
front is trapped by the sector of the shell which is evolving into the
extended halo.  It has been noted in \citet{gtt99} that P-Cygni profiles
are seen in galaxies that are on the higher luminosity side of the
distribution ($M<-18$) in their sample. Similarly, at high redshift the
Lyman Break Galaxies (hereafter LBG) exhibit similar profiles, as discussed
above. Hence
despite the much larger strength of the starbursts they are hosting, when
compared to local ones, the HI gas is still present between the HII region
and the observer.  In particular this implies that in all these cases the
ionizing radiation has to be trapped, and that a low density halo is
surrounding these objects - as discussed above, without this halo the shell
would have been disrupted and no P-Cygni profile would be produced.

It is interesting to enlight this result with the constraints on the Lyman
continuum (hereafter LyC) radiation from galaxies. Most of the previous
observations of galaxies below the Lyman break \citep{leith95a,hur97} as
well as recent measurements obtained from FUSE observations
\citep{dehar01,heck01b} show that the fraction of ionizing stellar photons
that escape the ISM of each galaxy is small, if any. Most estimates led to
an escape fraction $\leq 5 \%$. However these figures are in constrast with
a LyC escape fraction larger than 50\% reported by \citet{steidel01}
obtained from a composite spectrum of galaxies at $<z>$ = 3.4. Indeed
starburst galaxies at large redshift have a higher UV luminosity than most
local starbursts studied so far. This would imply more photons for ionizing
the gas and probably an easier escape. However this difference remains a
puzzle since one must reconcile two facts that seem to contradict each
other. One is the leaking of the UV LyC and the other is the presence of
P-Cygni \lyal\ profiles. The very presence of these P-Cygni \lyal\ profiles
testifies that a substantial amount of HI is present, and hence should act
as a screening agent against the escape of the LyC. One way out would be to
assume that the HI layers are very perturbed and chaotic, so that they do
not completely cover the massive stellar clusters. However the \lyal\
emission that would escape from regions devoided of neutral gas should be
overwhelming, resulting in a symmetric \lyal\ emission profile. We predict
therefore that LyC photons would be detected mostly on galaxies showing
strong and quite symmetric \lyal\ emission lines, rather than in objects
showing typical P-Cygni profiles.  As far as we know, no relation has yet
been established between the fraction of escaping LyC and the \lyal\
emission profile in galaxies hosting starbursts at any redshift.

There is a final point we want to remark. As discussed at the end of
  Sect.~4,
the detection
of \lyal\ P-Cygni favours short-lived (or nearly instantaneous) bursts
rather than steady state, long lasting star formation processes. 
 After long-lasting star formation episodes the expanding shell would have
  finally blown-out in the intergalactic medium, and no P-Cygni profiles
  would be expected on the \lyal\ lines. We would rather expect to detect
  pure emission or absorption, depending on the orientation of the line of
  sight with respect to the disk of the galaxy.
In the
case of short-lived starbursts, the usual concept of star formation rate
(amount of gas mass being transformed into stars per unit time) becomes
meaningless: in most starbursts, massive star formation has probably ceased
already few Myr ago. It might be therefore misleading to derive the star
formation rate density at high redshift from the analysis of galaxies which
are forming stars in short bursts. To solve this inconsistency we propose
to parameterize the strength of these star forming episodes by the total
amount of gas transformed into stars, rather than by the rate of star
formation, and to reevaluate accordingly the history of star formation in
the early Universe. This analysis is out of the scope of this work, and
will be presented in a forthcoming paper.

\section{Summary and conclusions}

Our findings have important implications for the study of \lyal\
profiles in star-forming galaxies in particular, and for super-bubble
evolution and galactic winds in general.  We find that the P-Cygny profiles
are extended, smooth and span several kiloparsecs covering a region much
larger than the starburst and comparable or larger than the host galaxy
itself. This strongly suggests the existence of an expanding super-shell
generated by the mass and energy loss of the starburst interacting with an
extended low density gaseous halo.

We identify six phases in the predicted evolution of a super-shell
that correspond to the different types of \lyal\ profiles observed.

\begin{itemize}

\item[i)] The early phase. In a very young starburst the recombination lines
will be at their maximum strength and \lyal\ will show a broad absorption
centered at the systemic velocity of the galaxy. SBS 0335 could be in this
stage.

\item[ii)] The emission phase. The super-shell breaks, ionizing radiation
escapes into the low density halo fully ionizing a ``bi-conical''
region. \lyal\ is in emission at the systemic velocity of the galaxy. 

\item[iii)] Late emission phase. Recombination starts to dominate in the
expanding shell producing an additional \lyal\ component shifted by
the shell expanding velocity. T1214 would be at the end of phase ii,
starting already this phase. 

\item[iv)] Shell recombination phase. Recombination dominates this phase
characterized by a strong P-Cygni type profile in \lyal. The absorption
component is blue shifted by $v_{exp}$, the expansion velocity of the
shell.  \lyal\ emission is strongly depleted and the centroid of its
profile is shifted toward the red.  Backscattering and emission from the
receding part of the shell may give raise to an extended red wing in the
observed \lyal\ profile.

\item[v)] If the leading shock becomes radiative an additional unabsorbed
and blue shifted component of \lyal\ may be produced.

\item[vi)] Late phase.  The shell is almost completely recombined and has
substantially slowed down its expansion. \lyal\ will show a broad
absorption with only a small blue shift.  The recombination lines from the
HII region will be very weak.

These 6 phases and the associated 4 different profiles are able to
describe, both qualitatively and quantitatively, the variety of observed
profiles in both the nearby HST sample and high redshift star--forming
galaxies. Given that the probability of occurrence of a given profile
depends mainly on the age of the starburst (and also on the orientation)
the observed fraction of systems in the different phases will provide
important information to critically test the predictions of this scenario.

\end{itemize}

 Our results stress the importance of the density and kinematical
  structure of the neutral gas surrounding an HII region on the
  detectability of the \lyal\ emission. The observed properties of the \lyal\
  emission line will be a convolution of several effects.  
  While the amount of dust alone determines only the absolute intensity of 
  the emission line, but not affecting its equivalent width, the kinematical
  configuration of the neutral gas is the driving factor for its
  final visibility, and its profile shape -- from broad absorption to pure
  emission. The fact that several independent effects play such a
  significant role on the properties of the observed \lyal\ emission lines
  explains the lack of correlations established in the past between the 
  \lyal\ emission strength and other properties of different starburst
  galaxies. The use of \lyal\ as a tracer of star formation rate, and even
  as a redshift indicator for galaxies at high-z, should be done with care,
  always being aware of its limitations. 

\acknowledgments

JMMH has been partially supported by Spanish grant AYA2001-3939-C03-02. ET 
has been partially supported by the Mexican Research Council (CONACYT)
grant 32186-E. She also gratefully acknowledges the hospitality of the IoA
in Cambridge.  This work was supported by grant GO-08302.01-97 from the
Space Telescope Science Institute, which is operated by the Association of
Universities for Research in Astronomy, Inc., under NASA contract
NAS5-26555.

\clearpage

\clearpage

\clearpage

\clearpage

\clearpage

\clearpage

\clearpage

\clearpage

\clearpage

\clearpage

\clearpage

\clearpage

\clearpage

\clearpage

\clearpage

\clearpage

\clearpage

\begin{figure}
\epsscale{0.5}
\plotone{f17.eps}
\caption{
Effect on the resulting profile of the additional \lyal\ emission
  components. The first contribution originates at the inner part of the
  receding shell by photoionization and/or backscattering of \lyal\ photons
  produced in the central HII region. The example has been computed for
  log(nHI) = 19.5 cm$^{-2}$ and \vexp\ = $300$ \kms. The intensity of
  this contribution has been assumed as 10\% the intensity of the main
  \lyal\ component originated in the HII region, as found by
  \citet{legrand97} for Haro~2.  This component should appear redshifted at
  the velocity of the shell, in this case \vexp = $+300$ \kms, and enhances
  the base of the red wing. The second contribution to \lyal\ originates at
  the ionized region in front of the neutral expanding shell, appearing
  thus at the velocity of the approaching shell, \vexp\ = $-300$ \kms. This
  component is not affected by neutral gas scattering, and should appear on
  top of the damped absorption profile. In the top panel we show the three
  contributions to \lyal. The thick line is the emission line produced in
  the central HII region. The lower panel shows with thick line the
  resulting profile. The thin lines show the assumed absorption profile 
  and the
  convolved profile of the intrinsic \lyal\ line alone. The effect of these
  two additional components on the total resulting profile is evident. }
\label{bc1}
\end{figure}

\clearpage

\begin{figure}
\epsscale{0.9} 
\plotone{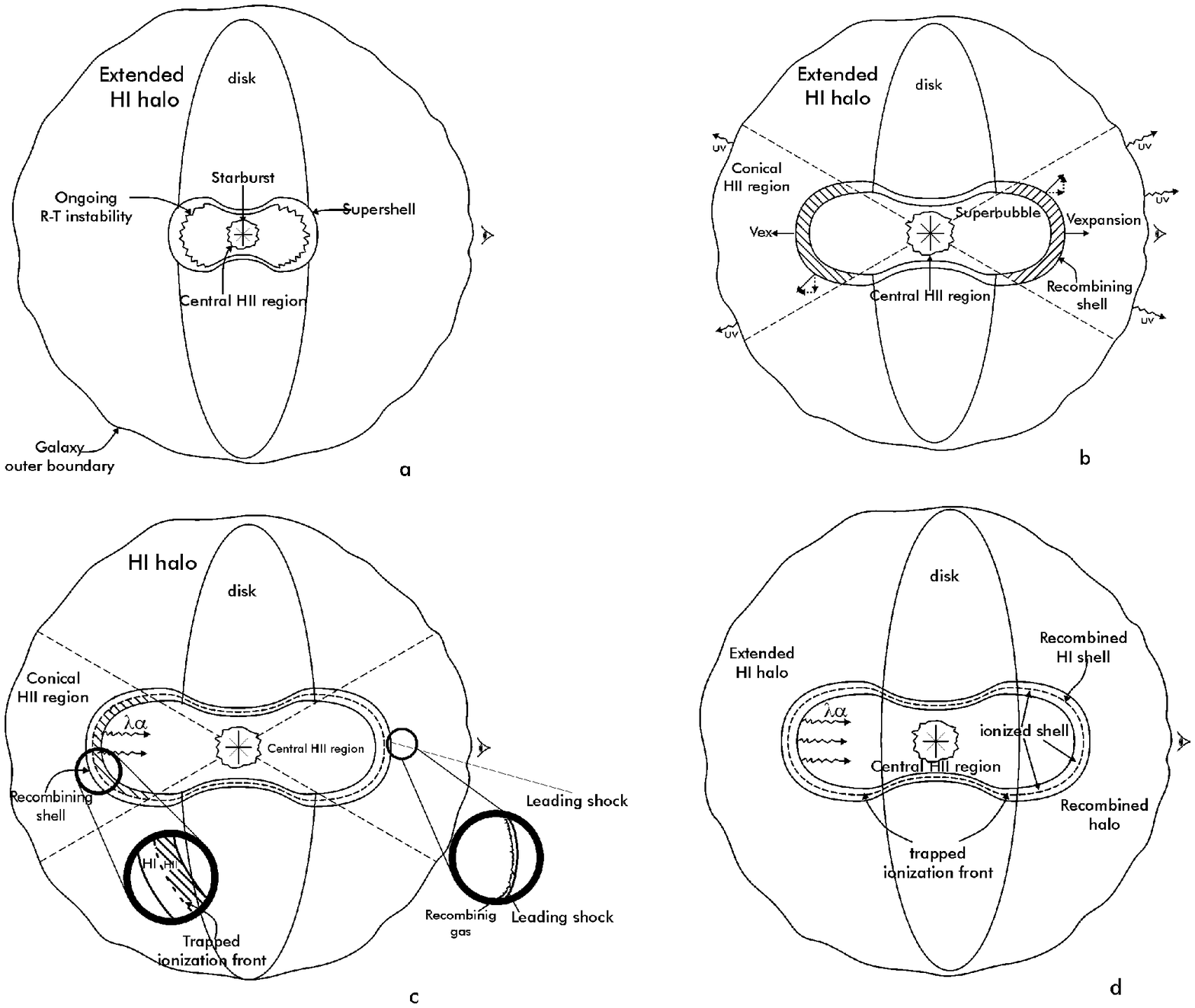}
\caption{
The basic model (see \citet{gtt99} for further details): evolution of the
expanding shell generated by an HII region and implications for the
visibility of the \lyal\ emission line.  a) A massive starburst generates a
central HII region. The surrounding halo of neutral gas absorbs all photons
with energy close to \lyal\ ones, producing a damped absorption profile.
b) At the beginning of the star-formation episode, the number of ionizing
photons emitted by the central cluster of massive stars is very large, so
that a fraction of them can escape the HII region and produce the
ionization of the surrounding halo of the host galaxy. An observer looking
straight through the ionization cone will detect a very strong \lyal\ line
centered at the rest velocity of the host galaxy. On the other hand, an
observer looking at a higher angle will still see  a damped absorption
profile produced by the neutral gas on the galaxy disk.  c) and d) The
action of stellar winds and supernovae explosions generates an expanding
shell that eventually will be able to undergo recombinations and  emits
also \lyal\ photons.  Furthermore, this large-scale expanding shell, driven
by the mechanical energy released by the massive central starburst, is
eventually able to trap the ionization front produced by the UV photons
that escape the central HII region.  This leads to the formation of two
zones within the shell, expanding at the same speed.  The inner zone is
fully ionized and emits in \lyal\ while the outer zone is neutral and
capable of scattering and absorbing this radiation. The neutral gas in the
approaching side of the shell leads to the formation of a \lyal\ P-Cygni
line profile, as discussed in the text. In addition, \lyal\ photons
produced by the central HII region and backscattered by the neutral layers
of the receding shell will contribute with a low intensity, broad component
redshifted by \vexp. }
\label{model1}
\end{figure}

\clearpage

\begin{figure}
\epsscale{0.7} 
\epsscale{0.7}
\plotone{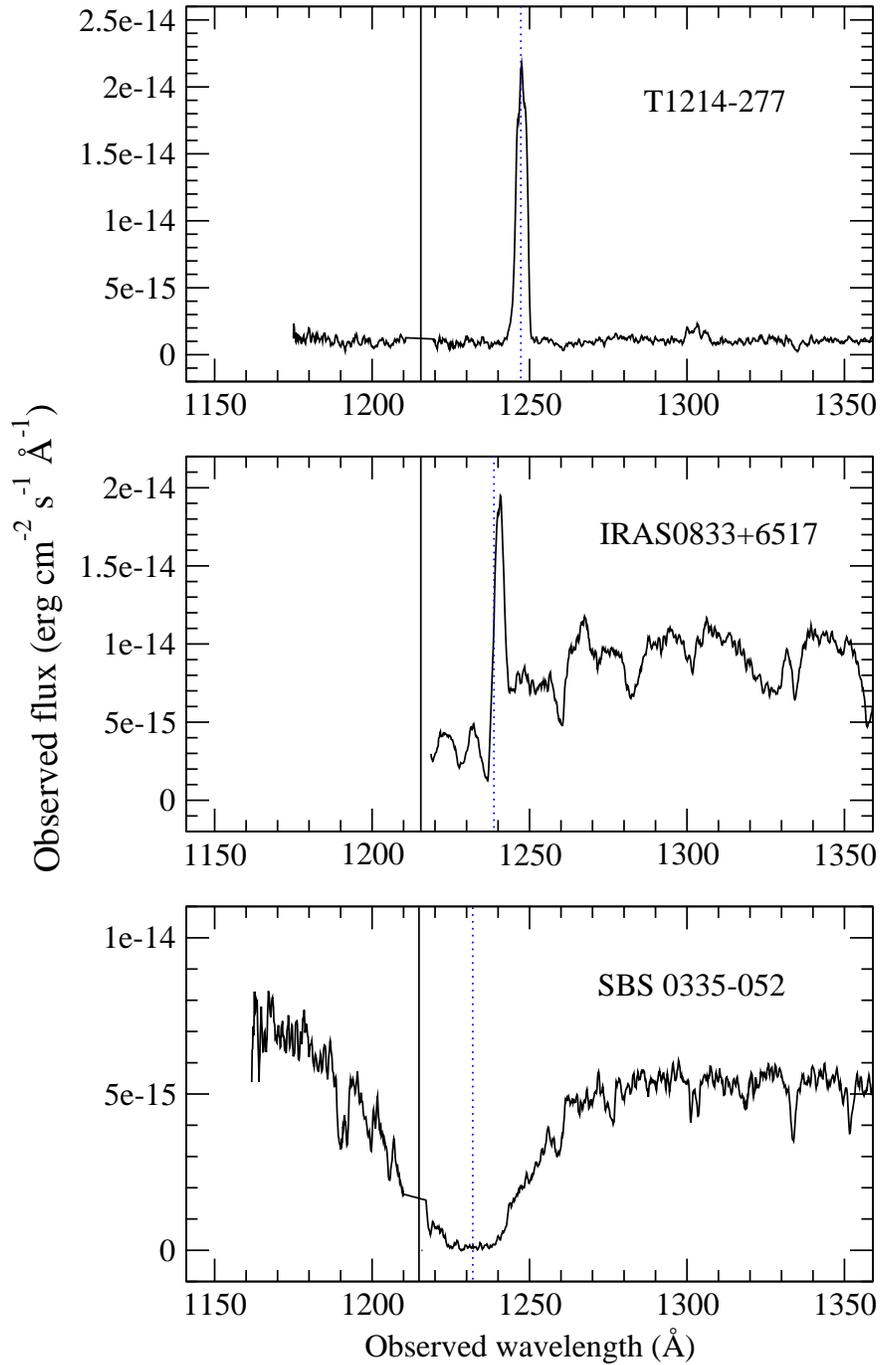}
\caption{Observed \lyal\  profiles illustrating the different cases
  discussed in the text: pure emission, with only weak absorption possibly
  produced by extinction; P-Cyg profile produced by a partially neutral
  expanding shell; and completely damped absorption produced by essentially
  static neutral gas.  }
\label{examples}
\end{figure}

\clearpage

\begin{figure}
\epsscale{0.7}
\plotone{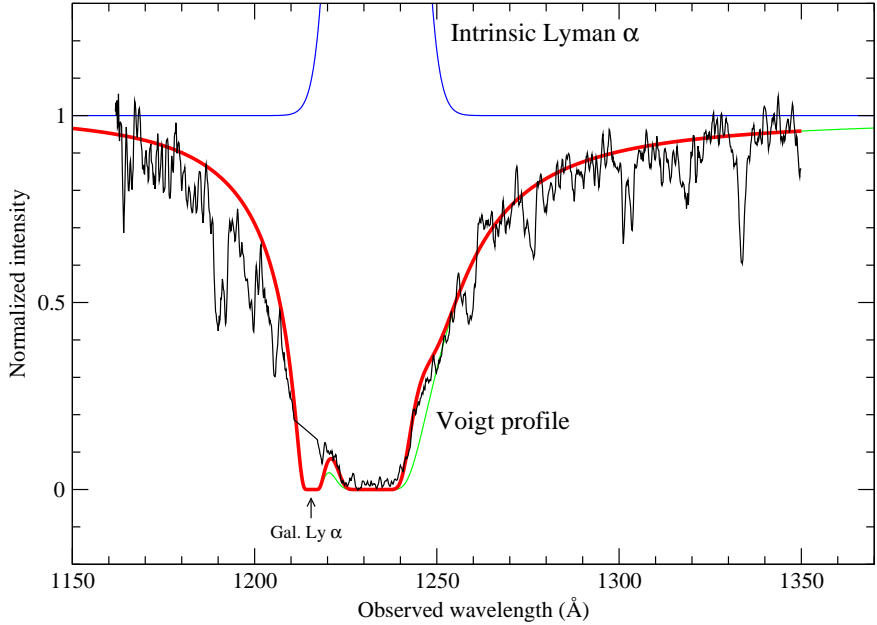}
\caption{Detail of the damped \lyal\ absorption profile in SBS
  0335-052. The thin line corresponds to the fitted Voigt absorption
  profile, including also the effect of the Galactic absorption.  The thick
  line shows the convolution of the intrinsic emission line and the total
  absorption profile.  It can be seen that the observed profile shows an
  excess with respect to the theoretical Voigt profile, detectable both on
  the red and the blue wings. We interpret this excess as the wings of the
  intrinsic \lyal\ emission line generated in the HII region, which become
  partially visible since the absorption at these wavelengths is not total,
  as evident from the Voigt profile.  The HII
  region \lyal\ emission required to reproduce the observations would have
  an equivalent width around 120 \AA, as expected for a very young starburst. }
\label{sbs1}
\end{figure}

\clearpage

\begin{figure}
\epsscale{0.7}
\plotone{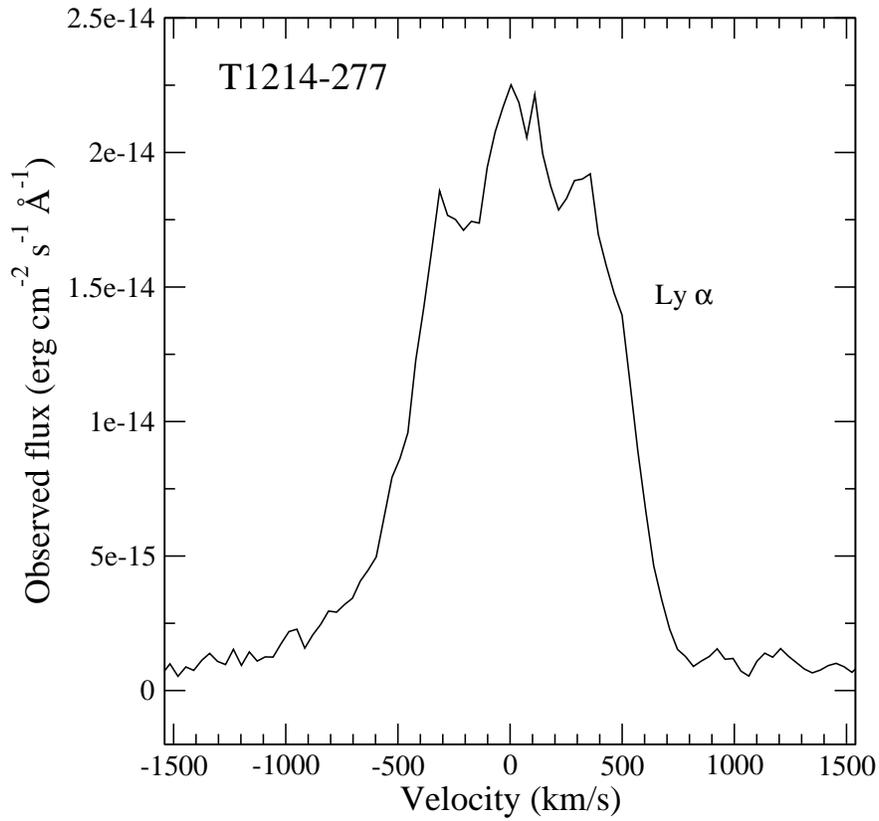}
\caption{Detail of the T1214-277 \lyal\  emission profile in velocity scale. 
The 2 peaks identified at $\sim \pm 300$ \kms\ from the central line peak
could be attributed to the secondary emission from the approaching and
receding parts of the young shell, according to steps ii -- iii in the text. 
}
\label{t1214-la}
\end{figure}

\clearpage

\begin{figure}
\epsscale{0.7}
\plotone{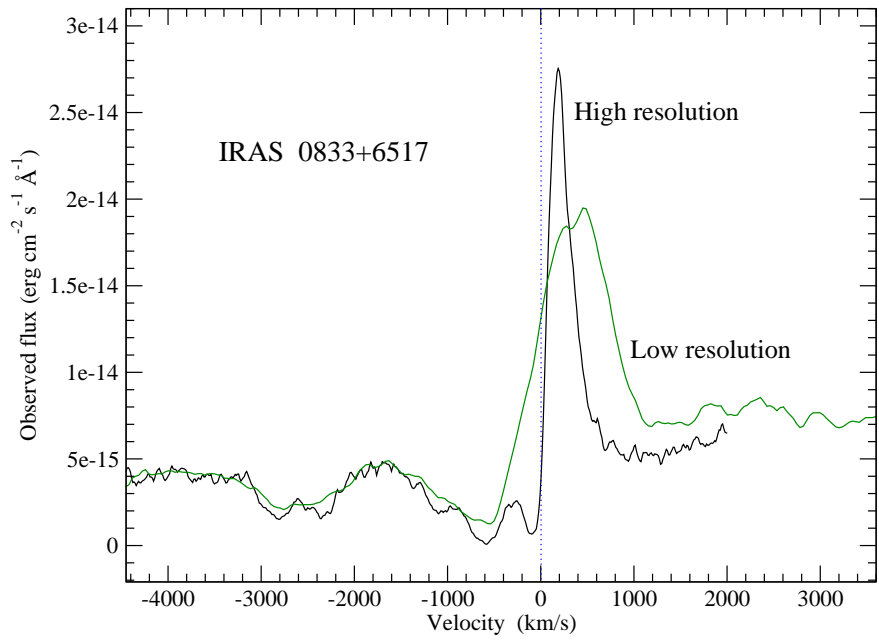}
\caption{Detail of the IRAS~0833+6517 \lyal\  profile observed at two 
  different resolutions.
  Note that low resolution spectroscopy can hide the presence of a
  blueshifted damped \lyal\  absorption profile. }
\label{iras08l-h}
\end{figure}

\clearpage

\begin{table}
\caption{Adopted properties of the observed HII Galaxies taken from the NASA
Extragalactic Database, except for the metallicity for which the references
are indicated.\label{sample1}}
\begin{tabular}{lccccccc}
\tableline\tableline
Object  & RA(2000)  & Dec(2000)  & M(B) & v(hel) &  Distance&Scale & 12+log(O/H)\\
        &           &            &      & \kms &    Mpc&pc/\arcsec &\\
\tableline
IRAS~0833+6517 & 08 38 23.2 & 65 07 15 & -20.8 & 5730 & 85.4&413& 7.5$^1$ \\
IZw18 & 09 34 02.4 & 55 14 32 & -14.0 & 780  &10&48.4& 7.2$^2$ \\
Haro~2  & 10 32 31.9 & 54 24 03 & -18.2 & 1461 &19.5&94.4& 8.4$^3$ \\
\tableline
\end{tabular}
\tablerefs{1- \citet{margon}; 2- \citet{skillman}; 3- \citet{davidge}} 
\end{table}

\clearpage 

\begin{deluxetable}{lccccc}
\rotate
\tablewidth{0pt}
\tablecaption{Journal of observations for the HST proposal 8302. The 
52$\times$0.5 slit was used in all cases. The UV spectra were obtained with
the FUV-MAMA detector, at a central wavelength of 1222 \AA. The optical
spectra were obtained with the CCD, at a central wavelength of 4300 \AA. 
The position angle of the Y axis is measured in degrees E of N.  The
coordinates indicated correspond to the center of the slit, as extracted
from the STIS files headers (J2000).
\label{observations1}}
\tablehead{
\colhead{Object}  &  \colhead{Obs. date} & \colhead{Grating}& 
\colhead{Integration time (s)}  &  \colhead{Pos. angle (deg)} &
\colhead{Coordinates  (RA, Dec -- deg)}}
\startdata
IRAS~0833+6517 & 2001-01-15 & G430L & 360 &  171 & 129.59666, 65.12083\\
IRAS~0833+6517 & 2001-01-15 & G140M & 1320, 3000, 3000 & 171& 129.59666, 65.12083 \\ 
IZw18 &2000-10-04& G430L&300&$-96$ & 143.50833,  55.24094\\
IZw18 &2000-10-04& G140M&1764, 3135, 3089, 3089& $-96$ & 143.50833,  55.24094\\
Haro~2 \#1\tablenotemark{a} (major axis)  & 2000-02-21 &G430L&300&150 & 
158.13250, 54.40972\\
Haro~2 \#1\tablenotemark{a}  & 2000-02-21 &G140M&1724, 3111, 3065&150 &
158.13250, 54.40972\\
Haro~2 \#2 (minor axis) & 2000-12-01 & G430L&300&$-121$ & 158.13250, 54.40097\\
Haro~2 \#2 & 2000-12-01 & G140M&1724, 3111, 3065&$-121$ & 158.13250, 54.40097\\
\enddata
\tablenotetext{a}{Due to an error, the center of the slit was misplaced 
  31\arcsec\ to the N during this visit.}
\end{deluxetable}

\clearpage

\begin{table}
\caption{Summary of observational results. The emission lines intensities
  have been measured through the same apertures in the UV and optical, so
  that their ratios reflect the intrinsic ratios in the gas. \lyal\ has
  been measured extrapolating the UV continuum from longer wavelengths. The
  fluxes are given in units of erg s$^{-1}$ cm$^{-2}$. 
\label{observations2}}
\begin{tabular}{lcc}
\tableline\tableline
&IRAS~0833+6517&Haro~2\\
\tableline
$F[OII]3727$&6.0e-14&1.2e-13\\
$F[OIII]4959$&2.3e-14&6.9e-14\\
$F[OIII]5009$&6.5e-14&2.0e-13\\
$F($\hbeta$)$&2.0e-14&7.7e-14\\
W(\hbeta) (\AA)&8&75\\
$F($\lyal$)$&4.3e-14&1.6e-13\\
W(\lyal) (\AA) &6&12\\
$F(sec.$\lyal$)$&4.0e-15&-\\
Aperture& 3\farcs5$\times$0\farcs5 & 1\farcs5$\times$0\farcs5\\
\tableline
\end{tabular}
\end{table}

\end{document}